\input ./style/arxiv-general.cfg
\documentclass[MSNbibl,nameyear,dvips]{arxstspdf}
\makeatletter
   \@ifpackageloaded{graphicx}{}{\usepackage{graphicx}}
\makeatother
\usepackage{dcolumn}
\usepackage{flushend}
\usepackage{stfloats}
\usepackage{shere}

\volume{30}
\issue{3}
\pubyear{2015}
\firstpage{297}
\lastpage{327}
\doi{10.1214/15-STS514}
\docsubty{FLA}

\makeatletter
\newcolumntype{d}[1]{D{.}{.}{#1}}
\fnbelowfloat
\makeatother

\begin{document}
\begin{frontmatter}

\title{Assessing the Potential Impact of~a~Nationwide Class-Based
Affirmative~Action~System}
\runtitle{Impact of Class-Based Affirmative Action}

\begin{aug}
\author[A]{\fnms{Alice}~\snm{Xiang}\corref{}\ead
[label=e1]{axiang@post.harvard.edu}}
\and
\author[A]{\fnms{Donald B.}~\snm{Rubin}\ead[label=e2]{rubin@stat.harvard.edu}}\vspace*{-14pt}
\runauthor{A. Xiang and D. B. Rubin}

\affiliation{Harvard University}

\address[A]{Alice Xiang is J.~D. Candidate at Yale Law School,
127 Wall St, New Haven, Connecticut 06511.  Donald B. Rubin is Professor of Statistics,
Department of Statistics, Harvard University, 1 Oxford Street, Cambridge, Massachusetts 02138,
USA (e-mail: \printead*{e1}; \printead*{e2}).}
\end{aug}

%
\begin{abstract}
{\fontsize{9}{11}\selectfont{We examine the possible consequences of a change in law school
admissions in the United States from an affirmative action system based
on race to one based on socioeconomic class. Using data from the
1991--1996 Law School Admission Council Bar Passage Study, students were
reassigned attendance by simulation to law school tiers by transferring
the affirmative action advantage for black students to students from
low socioeconomic backgrounds. The hypothetical academic outcomes for
the students were then multiply-imputed to quantify the uncertainty of
the resulting estimates. The analysis predicts dramatic decreases in
the numbers of black students in top law school tiers, suggesting that
class-based affirmative action is insufficient to maintain racial
diversity in prestigious law schools. Furthermore, there appear to be
no statistically significant changes in the graduation and bar passage
rates of students in any demographic group. The results thus provide
evidence that, other than increasing their representation in upper
tiers, current affirmative action policies relative to a
socioeconomic-based system neither substantially help nor harm minority
academic outcomes, contradicting the predictions of the ``mismatch''
hypothesis, which asserts otherwise.}}
\end{abstract}

%
\begin{keyword}
\kwd{Causal inference}
\kwd{multiple imputation}
\kwd{class-based affirmative action}
\kwd{racial affirmative action}
\kwd{law school admissions}\vspace*{-6pt}
\end{keyword}
\end{frontmatter}

\section{Introduction}

Affirmative action in higher education is one of the most contentious
social policies of recent decades in the United States, with polarized
views that intersect at the heart of modern American values of
diversity, meritocracy, and social justice. In the wake of the US
Supreme Court rulings on affirmative action in
\textit{Fisher v. University of Texas} (2013) and \textit{Schuette v. Coalition to Defend
Affirmative Action} (2014), understanding the effects of current
affirmative action policies relative to their possible alternatives is
especially relevant today. Although an extensive literature discusses
the role of fairness and legal precedence\vadjust{\goodbreak} in affirmative action, there
have been limited empirical studies examining the current system and
its alternatives. In particular, affirmative action in which students
receive preferential admissions based on their socioeconomic status
(SES) has been proposed as an alternative to racial affirmative action
(\citeauthor{Fallon}, \citeyear{Fallon};
\citeauthor{Kahlenberg}, \citeyear{Kahlenberg};
\citeauthor{Malamud}, \citeyear{Malamud}), with some studies examining the implementation of
SES-based affirmative action (hereafter abbreviated as SES AA) in a few
US states, yet little has been done to assess empirically what the
nationwide impact of such a change in policy would be. We use the
1991--1996 Law School Admission Council Bar Passage Study data to simulate
the outcomes of an SES AA\vadjust{\goodbreak} policy and evaluate its potential impact on
the demographic composition, graduation rates,  rates of attempting
the bar, and bar passage rates.\looseness=-1

\subsection{Existing Empirical Literature}

Most of the existing empirical literature on racial affirmative action
(racial AA) has suggested that it has a positive impact on racial
minorities, playing a vital role in placing minorities into more
selective schools and leading to better financial aid packages and
other advantages for minorities (\citeauthor{WightmanA}, \citeyear{WightmanA};
\citeauthor{WightmanB}, \citeyear{WightmanB}; \citeauthor{Epple}, \citeyear{Epple};
\citeauthor{ArcidiaconoA}, \citeyear{ArcidiaconoA}). In contrast, the ``mismatch
hypothesis'' has gained traction since the publication of \citet{Sander}, which contends that students enrolled in schools
where they have lower academic credentials than their peers due to
admission via
affirmative action tend to perform more poorly than they would have
performed in environments where they were better matched academically
to their peers. \citet{Sander} controversially concludes
that due to this mismatch in academic credentials, racial AA actually
\emph{hurts} black students in terms of their academic performance and
bar passage rates and thereby leads to \emph{fewer} black lawyers than
there would be without racial AA.

Sander's analysis has been challenged, however, by a number of scholars
for its mistakes in causal inference (\citeauthor{Ho}, \citeyear{Ho}; \citeauthor{Brief},
\citeyear{Brief}). Conclusions regarding the mismatch hypothesis have
been contradictory, with some, including \citet{Ayres}, actually finding some evidence that affirmative action
improves academic outcomes for black students due to a ``reverse
mismatch effect,'' whereby student performance improves due to help and
inspiration from their academically more advanced peers and professors.

To assess the effects of racial AA, the studies discussed above compare
the current system with a hypothetical counterfactual system without any AA at all,
which does not reflect the policy alternatives currently being debated.
Even if courts ruled against racial AA, it is likely that schools would
continue to implement policies that promote some form of diversity in
admissions. Some studies have examined this issue by leveraging state
data from Texas and California, which (in the late 1990s) banned racial
AA in their public university admissions and essentially implemented
SES AA. These studies have generally found dramatic declines of 30--50\%\vadjust{\goodbreak}
in the enrollment of underrepresented\vadjust{\goodbreak} racial minorities due to the bans
on racial AA as well as evidence for the mismatch effect (\citeauthor{Card}, \citeyear{Card}). According to their analyses, the ban on
racial AA led to improvements in graduation rates for minority
students, complicating the question whether racial minorities benefit
from racial AA (\cite{ArcidiaconoB}).

These studies have data from both racial and SES AA and consequently
should
lend insight into how schools and students respond to bans on racial
AA. Nevertheless, it is difficult to say whether their
conclusions should  be
generalizable to a \emph{nationwide} ban on racial AA. California and
Texas both have large minority populations and selective public
universities. Moreover, considering that their admissions systems still
showed strong racial preferences after the ban (\citeauthor{Long},
\citeyear{Long}), it is questionable whether their post-1990s systems
can truly be considered SES-based. Finally, although these studies find
significant decreases in the numbers of minorities enrolling in schools
without racial AA, this result might not hold as strongly if all
schools nationally adopted SES AA. With only a few states changing to
SES AA systems, minorities can opt to apply to universities that retain
racial AA in order to attend more selective schools. \citet{Card} find some evidence for this change in application
behavior among less qualified minority students but not among highly
qualified minority students. If a federal ban on racial AA were
instituted, however, there would be no advantage to minorities applying
to universities in other states.


\subsection{Overview: Simulating SES AA}

Given the lack of direct empirical evidence about the possible
nationwide impact of switching to SES AA, we simulate the changes in
enrollment across law school tiers (levels of prestige) and in student
academic outcomes (graduation and bar passage attempts and success
rates) under SES AA using data from the 1991--1996 Law School Admission
Council (LSAC) National Longitudinal Bar Passage Study (\citeauthor
{BPS}). Law school admissions are particularly appropriate for such a
simulation because they are more ``numbers-driven'' than admissions for
most other programs in higher education, depending heavily on
applicants' LSAT scores and undergraduate GPAs, thereby decreasing the
role of unobserved applicant factors, such as extracurricular
activities, personal statements, and letters of recommendation. Also,
because a standard goal of law school students is to pass the bar
exam, the bar passage rate provides a consistent metric for student success.\vadjust{\goodbreak}

Here we focus on the impact of a switch from racial AA to SES AA. We
take a potential outcomes, or Rubin Causal Model (\citeauthor{Holland}, \citeyear
{Holland}), approach to this problem. In particular, we consider two
possible treatment assignments being applied to admit students to law
schools: the first is the actual racial AA, and the second is a
counterfactual SES AA where the same numbers of low SES students are
admitted in each tier as there were black students admitted under the
racial AA. We have data on background characteristics and outcomes such
as tier attended, graduation, and bar attempts and passage under the
actual racial AA; our task is to predict what those outcomes would have
been under the alternative treatment, the counterfactual SES AA. These
predictions will combine explicit assumptions with relationships
between outcomes and covariates estimated from the racial AA data.
Because only two treatment assignments are being considered, all
students get subjected either to racial AA or to SES AA, and, as a
result, some common assumptions like the Stable Unit Treatment Value
Assumption (SUTVA, Rubin, \citeyear{RubinB}) are not relevant. All
assumptions are embedded within the imputation model being used to
predict the missing potential outcomes under SES AA. In this sense, our
framework is fully Bayesian.\looseness=-1

We model the current AA system by estimating separate ``tier enrollment
functions'' for black students and white students, where ``tiers''
capture the relative ranking of law schools. For the SES AA system,
students are reassigned to tiers by applying the black and white
student enrollment functions in each tier to the low and high SES
students, respectively, thus replacing race with SES as the selection
factor for AA. Based on these new tier assignments, the students'
graduation rates and bar passage rates were imputed. This process was
repeated forty times to multiply-impute the quantities of interest, as
recommended by \citet{Graham} for multiple
imputation of 50\% missing data (we have all of the results for racial
AA but are missing the results for SES AA). Thus, we were able to
compare the actual results of the current race-based system with the
counterfactual results of a hypothetical SES-based system to assess,
first, whether the latter would yield similar racial diversity across
tiers and, second, whether it would impact the graduation and bar
passage rates of students across demographic groups.

Our analysis addresses the mismatch effect where the source of mismatch
is discrepancies\vadjust{\goodbreak} in relative entering academic credentials (due to
racial AA), which is consistent with the definition of mismatch as used
in previous academic studies, but it does not address students'
feelings of mismatch stemming from being part of underrepresented
racial groups. Although there is reason to believe that diversity, in
terms of the proportion of black or low SES students in each
institution, would have an impact on minority performance, the
available data do not allow us to capture such effects in our model.
The data only specify the tier, not the particular institution, each
student attended, so the only data possibly relevant to diversity are
the proportions of minorities in the tiers. With only five tiers,
however, such analyses would be too crude to allow any meaningful
conclusions about the effects of diversity without making heroic and
unwarranted assumptions.

\section{The Data}

The BPS data were collected by LSAC from 1991 to 1996 from over 27,000
law school students, comprising 70\% of the entire incoming law school
class of 1991 in the US. Although it would be ideal to use more recent
data, unfortunately the study only spanned these years, and (as of this
writing) no comparable \mbox{nationwide} study with individual-level data has
been conducted since. The BPS includes the students' undergraduate GPA
(UGPA) and LSAT scores and the students' outcomes of law school tier
attended, law school graduation status, and bar passage status, all
obtained from the law schools and American Bar Association
jurisdictions. Also, all participating students were administered an
Entering Student Questionnaire that included self-reported race and
socioeconomic background. Although the questionnaire featured five
racial categories (white, black, Hispanic, Asian, and other), we
focused our analysis on white and black students because the data
revealed ambiguities regarding the extent to which the other racial
groups received preferential admissions under the current affirmative
action system (for further discussion, see Appendix~\ref{apprace}). We
used the version of this dataset prepared by \citet{Sander}.

\subsection{SES Categories}

The BPS does not include direct data on the family income of students,
but it does contain questionnaire responses from students about their
parents' occupations, education levels, and general socioeconomic
status, specifically, categories of occupation (from manual worker to
professional) and educational attainment (from high school dropout to
postgraduate degree) for both parents. Also, students ranked their family
income relative to American families in general with options ranging
from ``far below average'' to ``far above average.''

To assign students to SES categories, we first coded the responses for
parental characteristics and general SES on a scale from 1 to 5, with
larger numbers corresponding to higher SES. In cases where some SES
data were missing for a student, we imputed the missing values. We then
used the first principal component of the SES variable as our SES score
(for details of the methods used for the SES score, see Appendix~\ref
{appscore}). This principal component summarized 60\% of the variance
of these SES variables. We also assessed the sensitivity of our results
by using an alternative score.\footnote{$\mathit{SES\_Score} = \mathit{fam\_inc}^2 + \mathit{occ\_mom}
\cdot \mathit{ed\_mom} + \mathit{occ\_dad} \cdot \mathit{ed\_dad}$, where $\mathit{occ}$ is the
parent's occupation category, $\mathit{ed}$ is the parent's educational
attainment, and $\mathit{fam\_inc}$ is the student's response to the general SES
question. The $\mathit{occ}$ and $\mathit{ed}$ for each parent were multiplied to capture
the fact that the two factors carry complementary information.} None of
the results using the alternative score significantly differed and thus
are not reported here.

To establish an equivalence between the actual AA system and our
counterfactual SES one, we made the ``low SES'' group the same size as
the black student population and the ``high SES'' group the same size
as the white student population by using the corresponding SES score
percentiles and placing the students with lower SES scores into the
``low SES'' category. This mapping between racial groups and SES
categories ensures that the simulated SES AA system targets the same
number of students as the current AA system.

\subsection{Law School Tiers}

The study clustered the 163 participating law schools into six tiers,
with Tier~1 being the most selective and Tier~5 being the least
selective; Tier~6 was unique in that it consisted largely of
historically black law schools and had disproportionately large
representation from minorities. It is unclear how changes in AA
policies would impact Tier~6 schools. As shown in Table~\ref
{tier6lsat}, although the LSAT quartiles for white students in Tier~6
are slightly lower than those in Tier~5, the LSAT quartiles for black
students in Tier~6 are higher than those in Tier~5, which suggests that
Tier~6 is not less selective than Tier~5 and actually attracts more
qualified black students than Tier~5.\vadjust{\goodbreak} Tier~6 students appear to value
attending schools with larger minority populations, so changing to SES
AA seems irrelevant to Tier~6. Thus, we excluded Tier~6 and its
students from our analysis.

\section{The General Model for Law School Attendance}

In our simulation, we assume students will attend the highest tier
school to which they are admitted, and the number of students attending
each tier under SES AA is the same under racial AA. The relevant student
characteristics for admissions that are observed in the dataset are
LSAT, UGPA, race, and SES. In our model predicting the results of SES
AA, law schools switch from valuing racial diversity to valuing
socioeconomic diversity but do not change the extent to which they
value academic factors. To provide more structure to the model, each
law school tier is modeled as having a diversity quota, such that the
number of low SES students attending each tier under SES AA is the same as the
number of black students attending each tier under racial AA (see
Appendices~\ref{appquota} and \ref{appnoquota} for more details of the
diversity quota model).

We assume that the change from racial to SES AA would not lead to a
change in the population of law students nationwide. Although it is
possible that some students would decide not to attend law school at
all under SES AA and that others who did not actually attend law school
would under SES AA, we are unable to infer these results based on the
BPS data.

%
\begin{table}
\tabcolsep=0pt
\caption{LSAT score quartiles for Tiers 5 and 6}\label{tier6lsat}
\begin{tabular*}{\tablewidth}{@{\extracolsep{\fill}}@{}lcccc@{}}
\hline
& \textbf{Tier~5} & \textbf{Tier~6} & \textbf{Tier~5} & \textbf{Tier~6}\\
\textbf{Quartile} & \textbf{white} & \textbf{white} & \textbf{black} & \textbf{black}\\
\hline
25\% & 30 & 27 & 21 & 21 \\
50\% & 33 & 30 & 24 & 25 \\
75\% & 35 & 35 & 27 & 30 \\
\hline
\end{tabular*}
\tabnotetext[]{ta1}{\textit{Note}: Black students in Tier~6, comprising most of the Tier~6 population,
have higher scores than the black students in Tier~5, whereas white
students in Tier~5 have higher scores than the white students in Tier~6.
It is thus difficult to rank Tier~6 relative to the other tiers.}\vspace*{-5pt}
\end{table}

\subsection{Enrollment Probability Functions}

Because the BPS only includes data on enrollment and not on admissions,
instead of estimating each student's admissions probabilities to each
tier, we estimated each student's probability of enrollment into each
tier.\vadjust{\goodbreak} Specifically, we estimated the probability of a student being in
a given tier versus a lower ranked tier to obtain conditional tier
enrollment probability functions. To the extent that students enroll in
the best tier to which they are admitted, modeling affirmative action's
effects through the conditional enrollment probabilities is equivalent
to modeling them through admissions probabilities.

Ten separate enrollment functions were estimated, one for each of the
two racial groups in each of the five tiers. Within a racial group, a
student's tier enrollment probability was modeled as only depending on
the student's LSAT and UGPA. The conditional probability of student $i$
enrolling in tier $t$ was estimated using a sequence of logistic
regressions by first comparing those enrolled in Tier~1 versus those in
Tiers 2--5, followed by those in Tier~2 versus those in Tiers 3--5, and
so on, where, for student $i$ of race $r$ ($b$ for black or $w$ for white),
\[
p_{i,t}^r = \operatorname{logit}^{-1}\bigl(
\alpha_{t,0}^r+\alpha_{t,1}^r \cdot
\mathrm{LSAT}_i + \alpha_{t,2}^r \cdot \mathrm{UGPA}_i
\bigr).
\]
We estimated the logistic regressions using the\break \emph{bayesglm}
function in R with the default recommended prior distributions (\citeauthor{Gelman}, \citeyear{Gelman}).\footnote{We used Student-t prior
distributions with mean 0 and scale 2.5 for the coefficients. The prior
distribution for the constant term was set so it applied to the value
when all predictors are set to their mean values.}

%
\begin{table}
\tabcolsep=0pt
\caption{Regression coefficients for probabilities of enrolling in given tier versus lower tier}\label{regcoeff}
\begin{tabular*}{\tablewidth}{@{\extracolsep{\fill}}@{}lcd{2.8}d{3.9}d{2.2}@{}}
\hline
&& \multicolumn{1}{c}{\textbf{Black}} & \multicolumn{1}{c}{\textbf{White}} & \\
&& \multicolumn{1}{c}{\textbf{coefficients}} & \multicolumn{1}{c}{\textbf{coefficients}} & \multicolumn{1}{c@{}}{\textbf{T-statistic}}\\
\textbf{Tier} && \multicolumn{1}{c}{\textbf{(SE)}} & \multicolumn{1}{c}{\textbf{(SE)}} & \multicolumn{1}{c@{}}{\textbf{of difference}}\\
\hline
1 & Intercept & -12.95~(0.97) & -16.46~(0.33) & 3.43 \\
& LSAT & 0.20~(0.02) & 0.22~(0.006) & 1.39 \\
& UGPA & 1.47~(0.23) & 1.57~(0.07) & 0.41
\\[2pt]
2 & Intercept & -7.51~(0.67) & -8.17~(0.20) & 0.95 \\
& LSAT & 0.12~(0.01) & 0.12~(0.004) & 0.14 \\
& UGPA & 0.91~(0.17) & 0.73~(0.05) & 1.02
\\[2pt]
3 & Intercept & -3.11~(0.58) & -10.57~(0.21) & 12.04 \\
& LSAT & 0.09~(0.01) & 0.15~(0.004) & 4.45 \\
& UGPA & 0.20~(0.15) & 1.44~(0.04) & 7.80
\\[2pt]
4 & Intercept & -4.12~(1.04) & -4.72~(0.29) & 0.56 \\
& LSAT & 0.14~(0.02) & 0.14~(0.006) & 0.01 \\
& UGPA & 0.71~(0.28) & 0.48~(0.06) & 0.83 \\
\hline
\end{tabular*}
\tabnotetext[]{ta2}{\textit{Note}:
Coefficients from logistic regressions where the outcome variable is
whether the student was enrolled in the given tier or a lower tier. As
expected, the coefficients for LSAT and UGPA are positive and
significant. Larger intercepts confirm uniform boosts in enrollment
probabilities, whereas larger coefficients on LSAT and UGPA imply
greater increases in enrollment probability per increase in academic
qualifications. Given that LSAT scores are on a scale roughly 10 times
that of UGPA (11--48 vs. 1.0--4.0), it appears that LSAT generally
contributes more to the tier enrollment probabilities than UGPA.}\vspace*{-5pt}
\end{table}

The results from these regressions are shown in Table~\ref{regcoeff}.
For more details about the algorithm used to estimate these tier
enrollment probability functions, see\vadjust{\goodbreak} Appendix~\ref{appenroll}. The
admissions boost from racial AA is revealed for every tier by larger
values for $\alpha_{t,0}^b$ than for $\alpha_{t,0}^w$ (intercepts) for
all $t$, implying that black students have higher conditional
enrollment probabilities in each tier than white students with
equivalent academic credentials. The values of the LSAT and UGPA
coefficients for black and white students are relatively similar across
Tiers 1, 2, 4, and 5. All of the coefficients for Tier~3 significantly
differ between black and white students, though this seems to be driven
by the particularly large difference in intercept values.

\begin{figure*}
\includegraphics{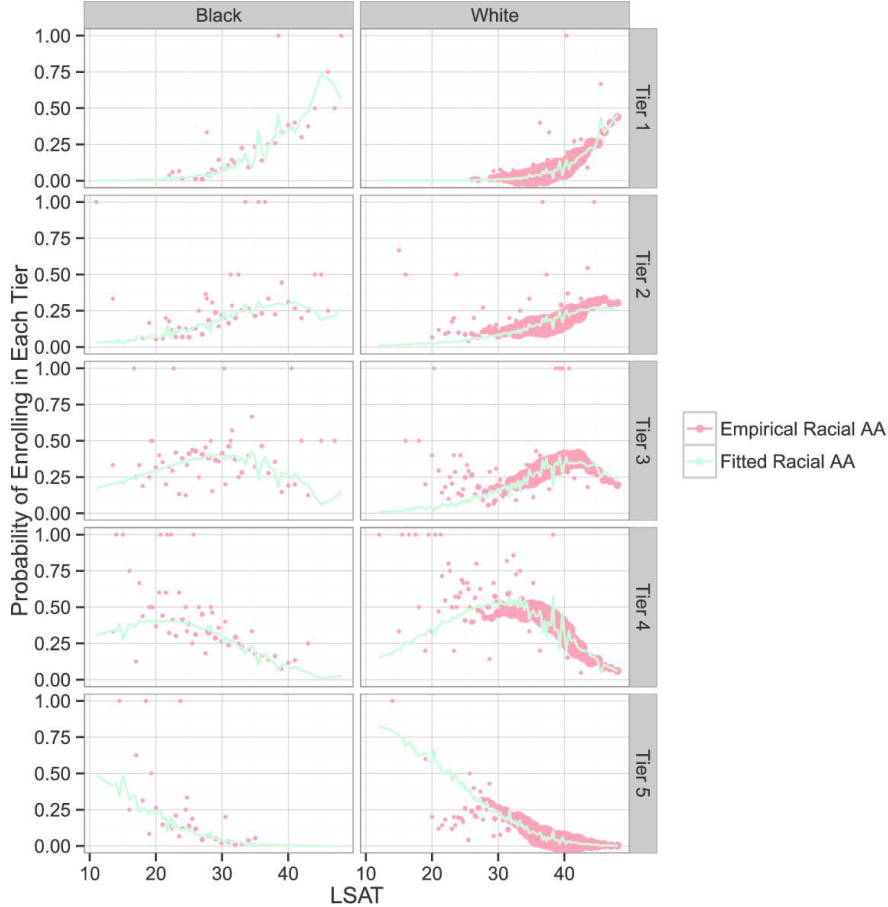}

\caption{Fit of model by race to current data.
Empirical proportions of students of each race with a given LSAT score
enrolled in a given tier (red dots), along with the fitted enrollment
probabilities for those students (green lines). The size of each dot
reflects the number of students with the corresponding LSAT score in
the tier. The fitted lines appear jagged because LSAT scores are not
continuous and because the functions depend not only on LSAT but also
on UGPA, a variable not displayed in these graphs.}\vspace*{-7pt}\label{figcheckrace}
\end{figure*}

\begin{figure*}
\includegraphics{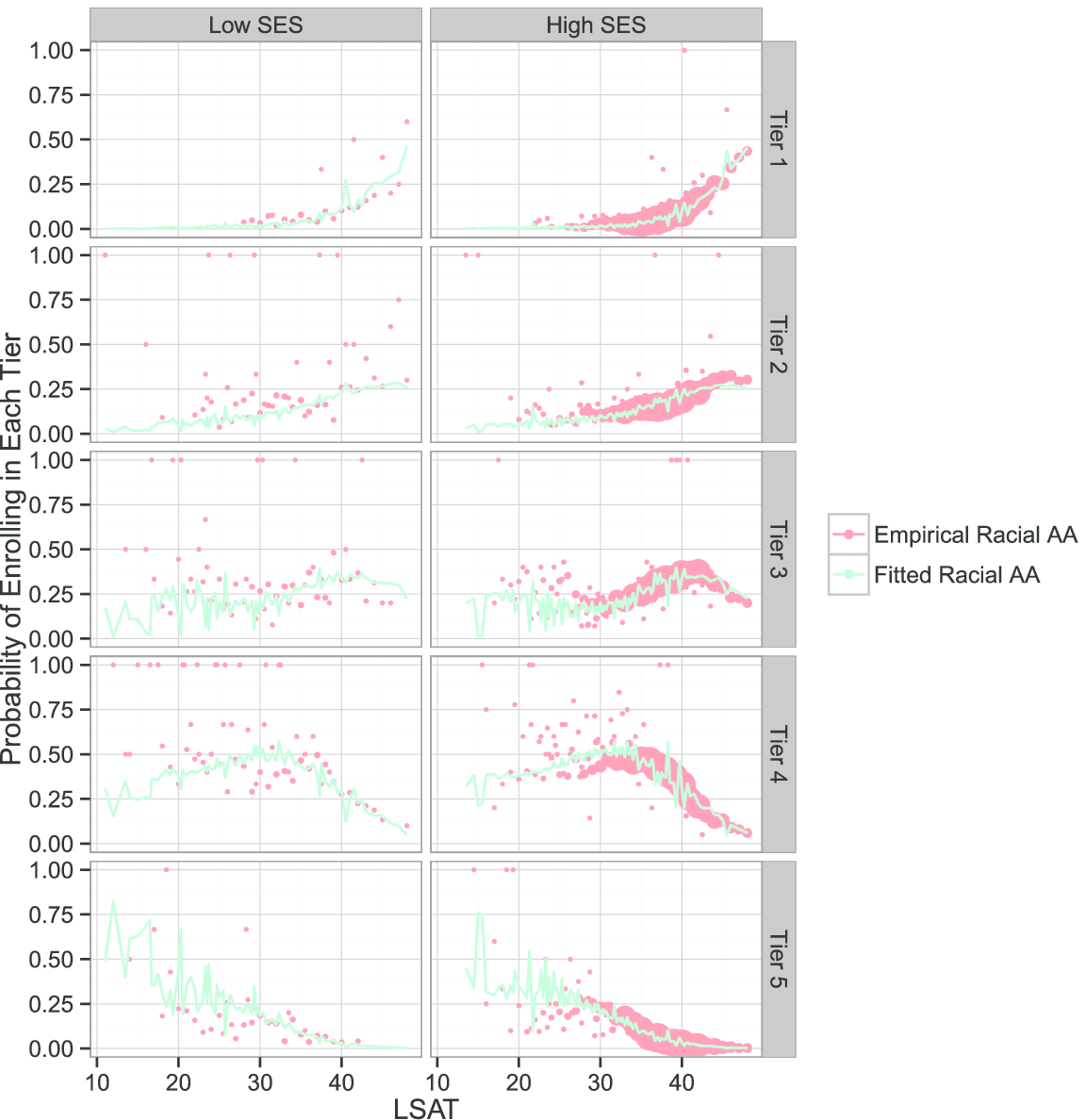}

\caption{Fit of model by SES to current data.
Empirical proportions of students of each SES group with a given LSAT
score enrolled in a given tier (red dots), along with the fitted
enrollment probabilities for those students (green lines). The size of
each dot reflects the number of students with the corresponding LSAT
score in the tier. The fitted lines appear jagged because LSAT scores
are not continuous and because the functions depend not only on LSAT
but also UGPA, which is not shown in the graphs. Given that the
estimated probabilities fit the data quite well, these graphs suggest
that there is no substantial discrepancy between the enrollment
functions of low vs. high SES students, supporting our assumption that
under the current system, only racial minorities benefit from
affirmative action.}\vspace*{-5pt}\label{figcheckses}
\end{figure*}

As shown in Figures~\ref{figcheckrace} and \ref{figcheckses}, the model
exhibits trends following the basic mechanisms of the existing AA
system: black students have boosted probabilities of being in higher
tiers. For example, the probability of a black student with a 40 on the
LSAT enrolling in Tier~1 is around 35\%, whereas it is only about 10\%
for white students with the same LSAT score. Figure~\ref{figcheckrace}
shows that for Tiers 1 and 2, the enrollment lines for the black
students are generally higher than those\vadjust{\goodbreak} for the white students, and
for Tiers 3--5, the peak of their enrollment functions are centered on
lower LSAT values. To further check the model fit, we used the model to
simulate the current AA system and to impute the academic outcomes. The
results, displayed in Appendix~\ref{appfit}, show that the model
accurately predicted all the quantities of interest.

%

%

\subsection{Reassigning Tiers}

\begin{figure*}
\includegraphics{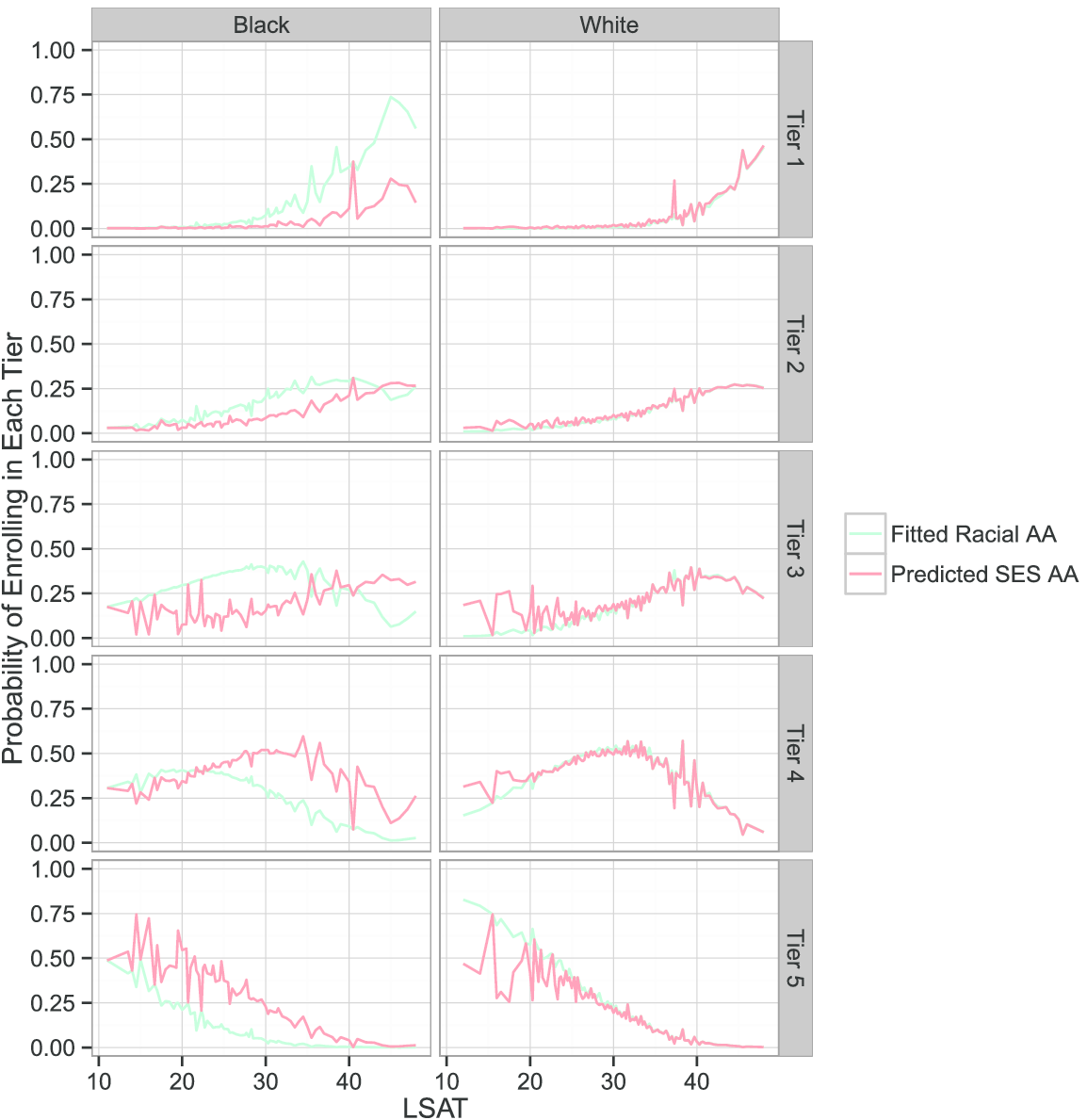}

\caption{Impact of SES AA on tier enrollment probabilities for students
by race.
Fitted probabilities for each student enrolling into each tier under
racial AA (green lines) and the estimated probabilities of enrollment
under SES AA (red lines). The lines appear jagged because LSAT scores
are not continuous and because the functions depend not only on LSAT
but also UGPA, which is not shown in these graphs.}\vspace*{-5pt}\label{figimpactrace}
\end{figure*}

\begin{figure*}\vspace*{-6pt}
\includegraphics{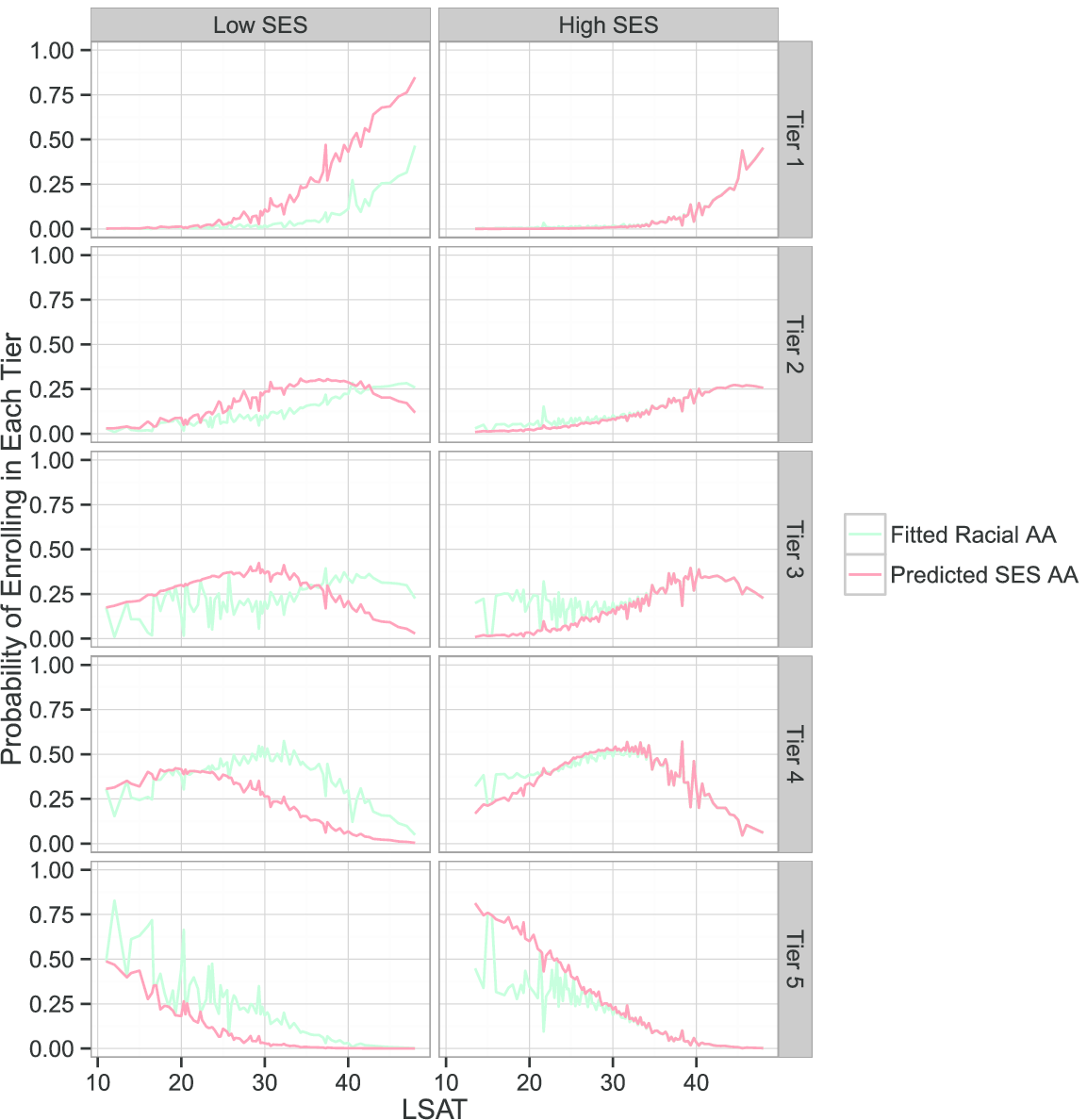}

\caption{Impact of SES AA on tier enrollment probabilities for students
by SES.
Fitted probabilities for each student enrolling into each tier under
racial AA (green lines) and the estimated probabilities of enrollment
under SES AA (red lines). The lines appear jagged because LSAT scores
are not continuous and because the functions depend not only on LSAT
but also UGPA, which is not shown in these graphs.}\vspace*{-10pt}\label{figimpactses}
\end{figure*}

To simulate SES AA, we assigned enrollment probabilities by considering
low SES and high SES students separately and applying, for each tier,
the estimated black student function to low SES students and the
estimated white student function to high SES students, with SES
indicators replacing the race indicators. The results are plotted in
Figures~\ref{figimpactrace} and \ref{figimpactses}, along with the
fitted curves from the original data under racial AA. Comparing these
curves shows the estimated impact of changing from racial to SES AA on
the students' probabilities of being enrolled in each tier. Starting
with Figure~\ref{figimpactrace}, under SES AA, the curves for the black
students now look similar to those for the white students. On the other
hand, Figure~\ref{figimpactses}\vadjust{\goodbreak} illustrates a significant boost for low
SES students under SES AA, comparable to
that given to black students,
meaning that the shapes of the curves for black and low SES students
essentially switched between the racial and SES AA systems.

The students were next assigned to tiers using these counterfactual SES
AA conditional enrollment probabilities. Students were first assigned
to Tier~1 by drawing from Bernoulli random variables with their
probabilities of enrolling in Tier~1. Once Tier~1 was full, Tier~2 was
next filled using the analogous procedure for the remaining applicants,
and so on until all students were assigned to tiers. The full algorithm
for assigning students to new tiers is described
in Appendix~\ref{appreassign}.

%

%

\subsection{Changes in Demographic Composition}

\begin{figure*}\vspace*{-7pt}
\includegraphics{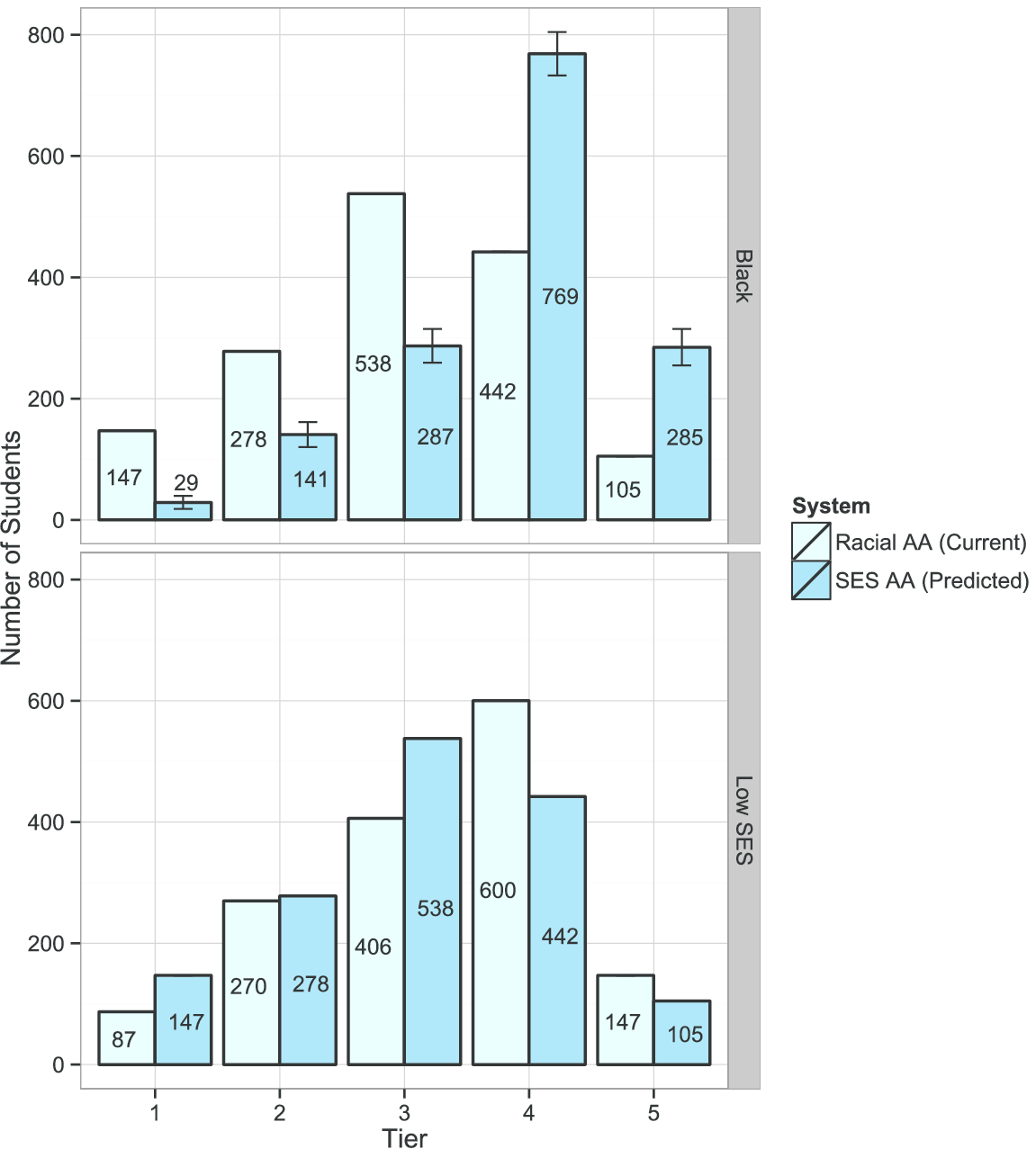}
\vspace*{-5pt}
\caption{Impact of SES AA on demographic composition across tiers.
Predicted changes in demographic composition by tier with SES AA. There
are predicted to be substantial decreases in the numbers of black
students in Tiers 1--3 and increases in Tiers 4--5, when switching from
racial AA (light blue) to SES AA (dark blue). The enrollment numbers
for low SES students under SES AA were fixed to equal those of black
students under racial AA. Generally, there are increases in the numbers
of low SES students in Tiers 1--3 and decreases in Tiers 4--5.}
\label{figdemcompchange}\vspace*{-12pt}
\end{figure*}

\begin{figure*}
\includegraphics{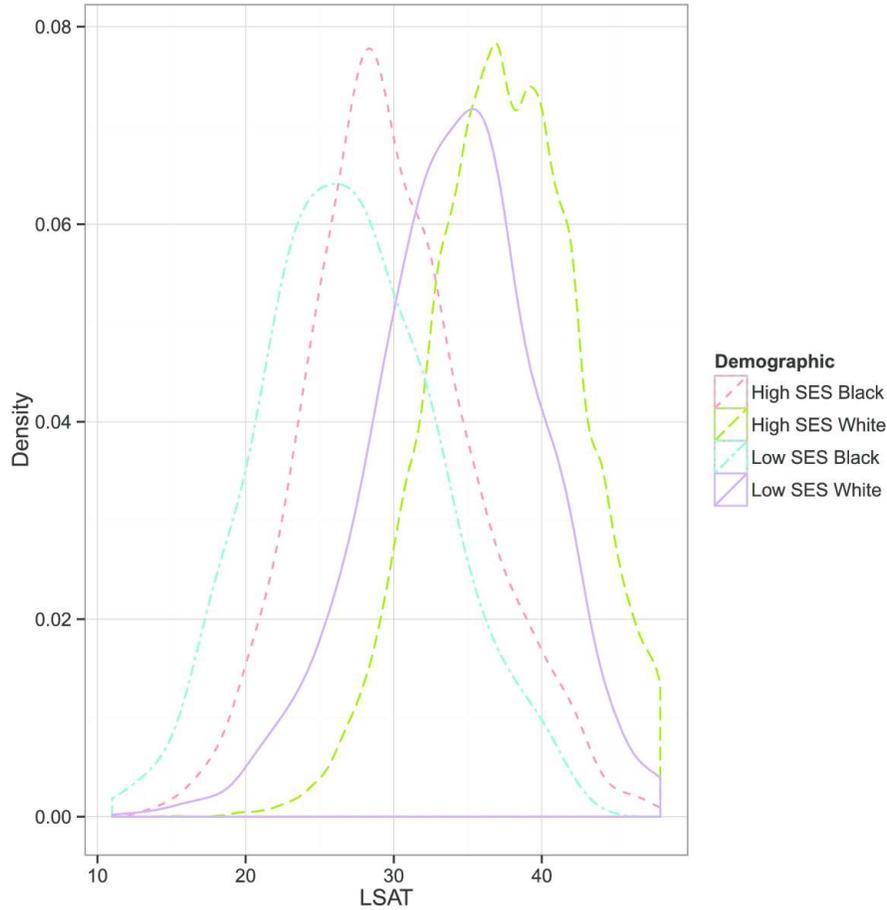}
\vspace*{-5pt}
\caption{LSAT distributions by intersected demographic group.
A much larger
difference exists between races within the same SES group than between
SES groups within the same race. The racial achievement gap is thus
larger than the SES achievement gap. When going from racial AA to SES
AA, low SES white students will thus benefit more than low
SES black students.}\label{lsatdistdemcrossfig}\vspace*{-8pt}
\end{figure*}

The results from the simulation predict substantial reductions in the
numbers of black students\vadjust{\goodbreak} in top tiers as a result of changing from
racial AA to SES AA: from 147 to an estimated 29 black students in Tier~1 and from 278 to an estimated 141 black students in Tier~2 (Figure~\ref
{figdemcompchange}). These dramatic changes stem from the fact that low
SES white students typically have higher LSAT scores than high SES
black students (see Figure~\ref{lsatdistdemcrossfig}), suggesting that
the switch from racial AA to SES AA replaces black students with low
SES white students. Even though low SES black students get the same AA
boost under either AA system, some of the black students currently admitted are
displaced by lower SES white students under SES AA.

\begin{figure*}\vspace*{-5pt}
\includegraphics{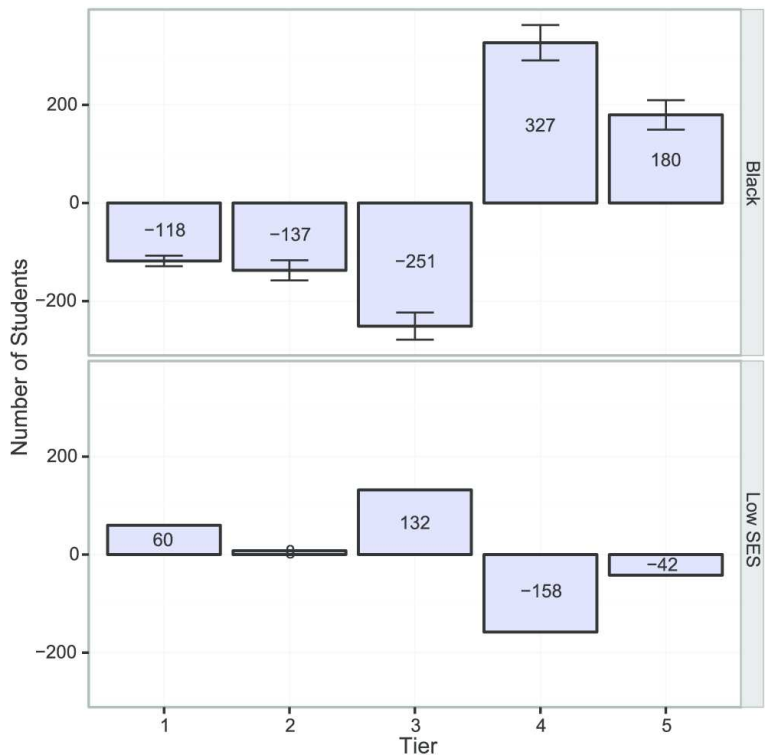}
\vspace*{-7pt}
\caption{Changes in demographic composition across tiers.
Predicted effect on black and low SES students when switching to SES AA
from racial AA. The result is a decrease in black students in Tiers 1--3
that is far greater than the increase in low SES students in these
tiers, a consequence of the larger achievement gap, as measured by LSAT
and UGPA, between black and white students than between low SES and
high SES students.}\label{figdemcompdiff}\vspace*{-10pt}
\end{figure*}

\begin{figure*}\vspace*{-3pt}
\includegraphics{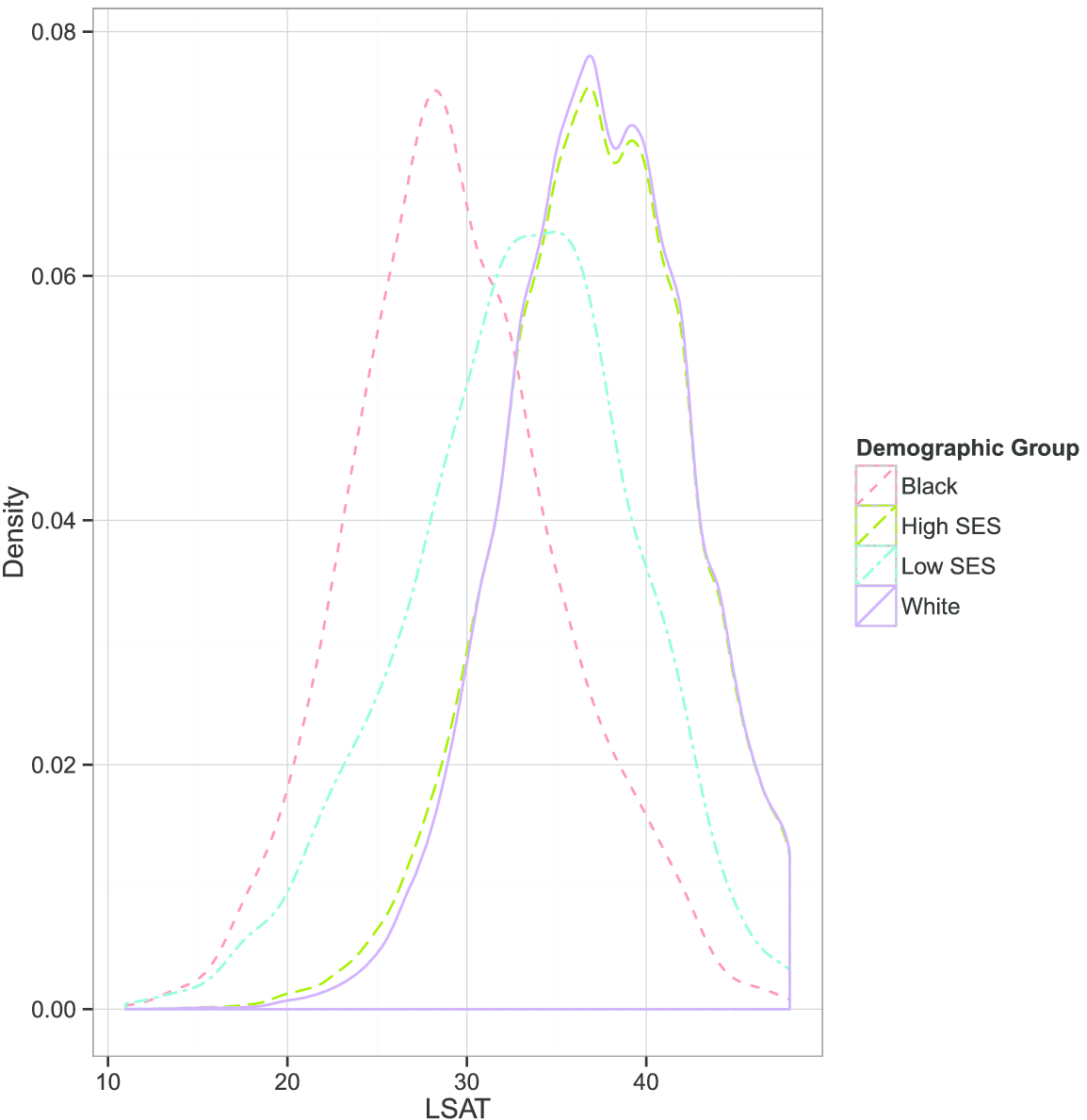}
\vspace*{-5pt}
\caption{LSAT distributions by demographic group.
LSAT distribution by demographic group. Black and low SES students have
lower LSAT distributions than white or high SES students, who have
essentially the same distributions. Notably, there is a substantial gap
between the black and low SES student distributions, providing evidence
that the LSAT gap between black and white students exceeds that between
low and high SES students.}\label{lsatdistdemfig}\vspace*{-8pt}
\end{figure*}

Moreover, under the SES AA system, the total estimated decrease in the
number of black students in Tiers 1--3 (506) substantially exceeds the
increase in the number of low SES students predicted to be admitted to
Tiers 1--3 (200), as illustrated in Figure~\ref{figdemcompdiff}, which
can be attributed to the fact that the achievement gap (i.e.,
differences in LSAT distributions) between black and white students
exceeds the gap between low and high SES students (Figure~\ref{lsatdistdemfig}). Thus, when low SES students are given the AA boost
rather than black students, the low SES students do not benefit as
dramatically as the black students did under racial AA. Under SES AA,
low SES students principally benefit from an increase in representation
in Tier~1 (from 87 to an estimated 147). Their numbers in Tier~2 are
virtually unchanged (from 270 to an estimated 278), and they only see a
moderate increase in representation in Tier~3 (from 406 to an estimated
538) and moderate decreases in Tiers 4 and 5 (from 600 to an estimated
442, and from 147 to an estimated 105, resp.), in contrast with the
dramatic tier composition changes experienced by black students under
the two systems.

It is noteworthy, however, that the sizable increases in the numbers of
low SES students in Tiers 1 and 3 did little to mitigate the
significant declines in the numbers of black students in those tiers,
which indicates that although SES and race are correlated (17\% of
black students are low SES, compared with 5\% of white students), there
is insufficient overlap to allow SES to serve as an effective proxy for
race: only 251 students are black \emph{and} low SES, out of 1510 black
students and 1510 low SES students.
This presents a significant policy
issue, because it suggests that it would be difficult to achieve racial
diversity without AA policies specifically targeted to admit black
students who are not low SES.

%

%
%
%

%
\begin{figure*}
\includegraphics{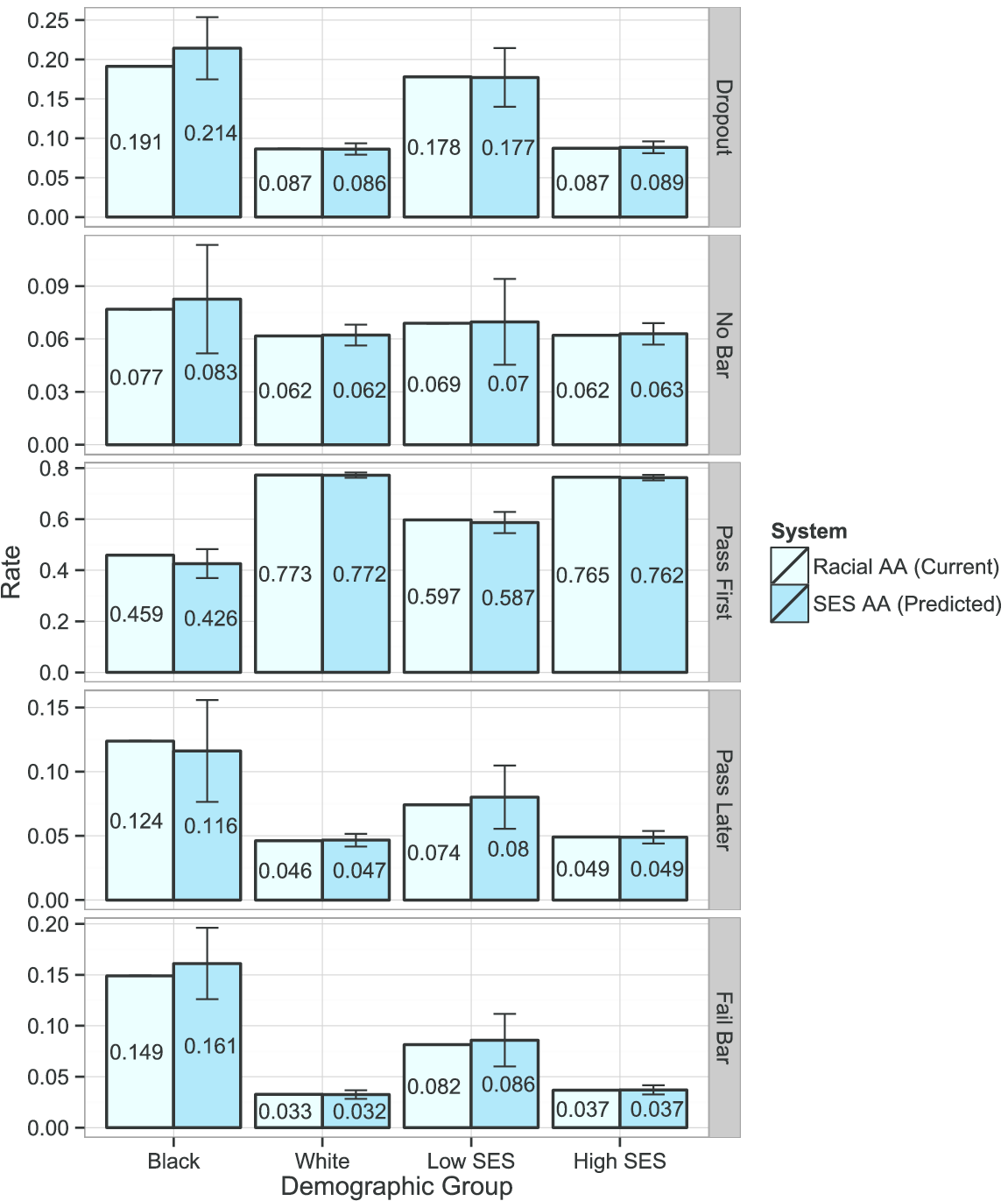}

\caption{Impact of SES AA on graduation and bar passage outcomes for
students by demographic group.
Neither graduation nor bar passage rates across demographic groups are
predicted to differ under racial AA (light blue) and SES AA (dark
blue). Note that although the academic outcomes were sequentially
imputed conditional on the previous step (e.g., we only predict whether
a student will pass the bar on their first try if we have imputed that
the student will attempt the bar), the results reported are the
unconditional rates for ease of interpretation. Given that each student
can only fall into one of the categories for outcomes, the rates within
each demographic group sum to~1.}\label{masterfig}
\end{figure*}
%

\section{Academic Outcomes}

In addition to demographic composition, we estimated the predicted
academic outcomes under the SES\vadjust{\goodbreak} AA system, simulating whether each
student would:
\begin{longlist}[iii.]
\item[i.] Graduate from law school,
\item[ii.] Attempt the bar exam,
\item[iii.] Pass the bar the first time,
\item[iv.] Pass the bar on a later try, or
\item[v.] Fail the bar.
\end{longlist}

We assumed that dropping out of law school implied not attempting the
bar, and thus not passing the bar.

\subsection{Imputing Graduation and Bar Passage Outcomes under SES AA}

The graduation and bar passage outcomes were imputed using a series of
logistic regressions. First, whether the students graduated\vadjust{\goodbreak} law school
was imputed by fitting a logistic regression to the current data's
dropout outcomes, using sex, LSAT score, LSAT percentile within tier,
race, and SES as predictors. Separate functions were estimated for each
tier so that imputing the students' new academic outcomes simply
involved applying the function corresponding to their SES AA tier assignment:
\begin{eqnarray*}
d_{i,t} &=& \operatorname{logit}^{-1}\bigl(
\alpha_{0}^t+\alpha_{1}^t\cdot
\mathrm{female}_i+\alpha_{2}^t \cdot
\mathrm{LSAT}_i
\\
&&\hspace*{33pt}{}
+\alpha_{3}^t \cdot \mathrm{LSATperc}_i
+\alpha_{4}^t \cdot \mathrm{UGPA}_i
\\
&&\hspace*{33pt}{} + \alpha_{5}^t\cdot \mathrm{LSATperc}_i \cdot
\mathrm{black}_i
\\
&&\hspace*{33pt}{} + \alpha_{6}^t\cdot \mathrm{LSATperc} \cdot
\mathrm{lowSES}_i
\\
&&\hspace*{15pt}\hspace*{33pt}{} + \alpha_{7}^t\cdot \mathrm{black}_i +
\alpha_{8}^t\cdot \mathrm{lowSES}_i\bigr),
\end{eqnarray*}
where $d_{i_t}$ is the probability of student $i$ in Tier~$t$ not
graduating (i.e., dropping out of) law school.

We included both LSAT score and LSAT percentile in order to better
detect any potential mismatch effect. Under the mismatch hypothesis, we
would expect that LSAT percentile might have a significant negative
effect on the chances of undesirable outcomes (dropping out and failing
the bar) and a\vadjust{\goodbreak} significant positive effect on desirable outcomes
(passing the bar), because students with lower LSAT percentiles within
their tier would perform more poorly even given identical LSAT scores.
What we find, however, is that almost all of the coefficients on LSAT percentile and its interaction with the black indicator variable are
insignificant (see regression coefficients in Appendix~\ref{appreg}).
Moreover, the few significant coefficients and most of the
nonsignificant coefficients have the opposite signs from what would be
expected under mismatch.

For each of the bar passage outcomes, the same methodology was
employed, removing students in each subsequent step once their outcomes
had been imputed\vadjust{\goodbreak} (the complete algorithm is described in Appendix~\ref
{appacout}). In order to attain estimates of the uncertainty for these
predictions, multiple imputation (\citeauthor{RubinA}, \citeyear{RubinA}) was used by
repeating the entire procedure forty times.


\section{Results for Academic Outcomes}

The results show that there are no substantial changes overall for the
student academic outcomes in going from racial AA to SES AA. Figure~\ref
{masterfig}, which summarizes the academic outcomes for each
demographic group under both AA systems, shows that all of the results
for the SES system are predicted to be within 95\% intervals of the
outcomes under the current racial AA system. Moreover, the magnitudes
of these effects appear to be mixed and minimal in the aggregate,
suggesting  either that the mechanisms behind them are limited in
effect or that they operate in opposite directions and cancel each other out. The
results by demographic group are shown\vadjust{\goodbreak} in Appendix~\ref{appdem} and by
tier in Appendix~\ref{apptier}.



Although the primary purpose of our simulation was not to estimate
mismatch effects, the changes in minority students' relative academic
credentials under SES AA in comparison to racial AA should allow us to detect
mismatch. Based on the figures in Appendix~\ref{applsat}, the overall
LSAT distributions for Tiers 1--3, the tiers where the mismatch effect
should be apparent, remained the same, although the LSAT distributions for
black students within each tier were shifted toward higher scores and
the LSAT distributions for low SES students were shifted toward lower
scores. Thus, black students under SES AA were more academically
qualified within each tier than they were under racial AA, and low SES
students were less academically qualified. Under these circumstances,
the mismatch effect predicts that the low SES students should have worse
academic outcomes and the black students should have better academic
outcomes in going from racial to SES~AA.

Given that bar passage outcomes generally did not improve for black
students or worsen for low SES students under the SES AA system, these
results suggest that incoming student characteristics are more
important in shaping academic outcomes than the tier boosts conferred
by affirmative action. Although differences in the regression
coefficients across tiers (Tables~\ref{regdropout}--\ref{reglater})
indicate that the specific tier a given student is in may have a slight
impact on academic outcomes, these impacts do not yield substantial
changes in aggregate performance.

%

\subsection{Dropout Rates}

For dropout rates, our simulation predicts no substantial changes
overall or on a\vadjust{\goodbreak} tier-by-tier basis (see Figure~\ref{dropouttierfig} in
Appendix~\ref{apptier}). Nonetheless, if we examine the direction of
the changes by tier, we see that predicted dropout rates for black
students increased for Tier~1 (from 4.8\% to 9.1\%, though with very
large standard errors), stayed roughly the same for Tiers 2 and 3, and
decreased for Tiers 4 and 5 (from 23.1\% to 21.3\%, resp.). For low SES
students, predicted dropout rates slightly increased for Tier~1 (from
6.9\% to 8.8\%), stayed virtually the same for Tiers 2--4, and increased
for Tier~5 (from 27.9\% to 30.6\%). Thus, with the exception of the
Tier~1 prediction for black students, on a tier-by-tier level, black
students have slightly better predicted dropout rates under SES AA,
whereas low SES students have slightly worse predicted dropout
rates.\footnote{Note that only seven black students in Tier~1 dropped
out in the actual data, so the prediction for that outcome has very
large standard errors and should be interpreted cautiously.}



\subsection{Rates of Not Taking the Bar}
For rates of not taking the bar exam, there are no predicted
significant changes overall and no apparent trends between tiers from
switching to SES AA from racial AA (Figure~\ref{nobartierfig} in
Appendix~\ref{apptier}).\vadjust{\goodbreak} For example, for black students, there is a
decrease in the proportion taking the bar in Tier~1 (4.2\%), increases
for Tier~2 (0.2\%) and Tier~3 (0.8\%), and decreases for Tier~4 (0.4\%)
and Tier~5 (0.6\%). This lack of a consistent pattern across tiers is
understandable given that there are many factors influencing a
student's decision not to take the bar exam. Some students might not
take the bar if they find better nonlegal job opportunities, whereas
others might not take the bar if they are worried about their ability
to pass.

\subsection{Bar Passage Rates}

As shown in Figure~\ref{masterfig}, black students are predicted to
have slightly lower
first-time bar passage rates under SES AA than under racial AA (45.9\%
to 42.6\%), and low SES students are predicted to have virtually the
same rates (59.7\% to 58.7\%). Black students also are predicted to
have slightly lower rates of passing the bar in a later attempt (from
12.4\% to 11.6\%) and higher rates of failing the bar (from 14.9\% to
16.1\%) under SES AA than under racial AA. Low SES students, on the
other hand, are predicted to have a slightly higher rate of passing the
bar in a later attempt (from 7.4\% to 8.0\%), but this effect does not
fully offset the decrease in low SES students predicted to pass the bar
on their first try, leaving the bar failure rate for low SES students
roughly the same (from 8.2\% to~8.6\%).


Although these changes are all within the 95\% intervals of the actual
values, they are notable in that they consistently contradict the
predictions of the mismatch hypothesis. The overall bar passage rate of
the low SES students was virtually unchanged under the SES AA system,
whereas the black students, no longer targeted by AA, had worse outcomes.

\section{Conclusion}
Our results provide some insight into the potential effects of adopting
SES AA, finding that (1) racial and SES AA achieve dramatically
different racial composition results, and (2) the data and our
simulations contradict the predictions of the mismatch hypothesis. In
particular, without affirmative action specifically targeting black
students,\vadjust{\goodbreak} attaining racial diversity in top law school tiers would be
very difficult. Although it is often argued that SES can serve as a
proxy for race, the data suggest that adopting an SES-based system would
not maintain racial diversity. Differences in applicant qualifications
across race persist even after controlling for SES, so most minority
students who are currently admitted and enroll into top schools are
from comparatively affluent backgrounds.

Moreover, assessing the impact of going from racial to SES AA on
student academic outcomes (graduation rates and bar passage rates)
revealed almost no significant differences between the results for the
two systems, even when examining the results under cross-sections of
race, SES, and law school tier. These results suggest that the
principal impact of affirmative action is on the racial and SES
composition across law school tiers rather than on academic outcomes
across racial and SES groups, a conclusion that contrasts with the
predictions of the mismatch hypothesis.\vadjust{\goodbreak}

The results show that affirmative action does not appear to have
negative effects on minority academic outcomes, but they also show
that, conditional on students' incoming academic credentials and
demographic characteristics, affirmative action does not appear to have
significantly positive effects either. Thus, from a policy perspective,
this analysis supports the need for racial affirmative action to
maintain racial diversity in upper law school tiers but also indicates
that improvements in minority academic outcomes would need to stem from
alleviation of the achievement gap in students' academic preparation
before law school.\vadjust{\goodbreak}

\begin{appendix}
\section{Model Selection}\label{appmodel}

\renewcommand{\thefigure}{\arabic{figure}}
\setcounter{figure}{9}
\renewcommand{\thetable}{\arabic{table}}
\setcounter{table}{2}

\subsection{Racial Groups}\label{apprace}
\begin{figure*}
\includegraphics{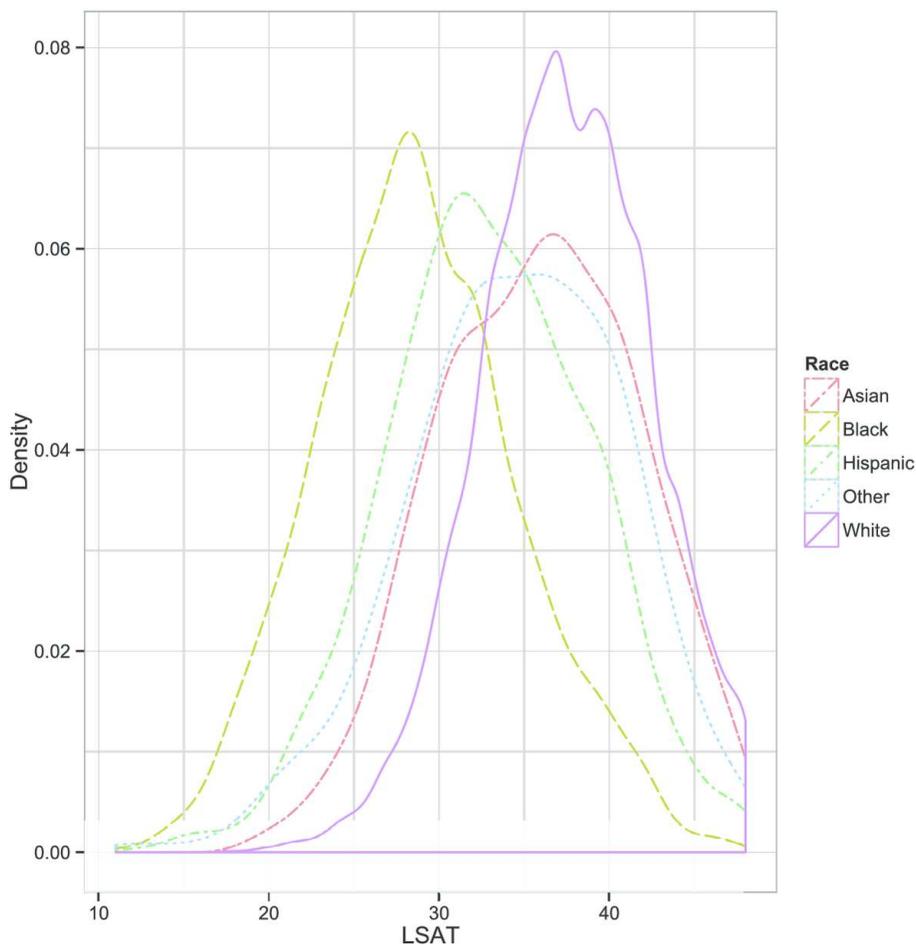}

\caption{LSAT distributions by racial group.
Although there are clear and substantial discrepancies between the LSAT
distributions for white and black students, the differences are smaller
for other racial minorities. In particular, Asians and ``Other''
students have similar distributions to white students, while Hispanics
are distributed between white and black students. This might explain
why the affirmative action trends were less apparent in the regression
coefficients for Hispanic, Asian, and Other students
in Table~\protect\ref{racecoeff}.}\label{lsatrace}\vspace*{-8pt}
\end{figure*}

\begin{figure*}
\vspace*{15pt}
\includegraphics{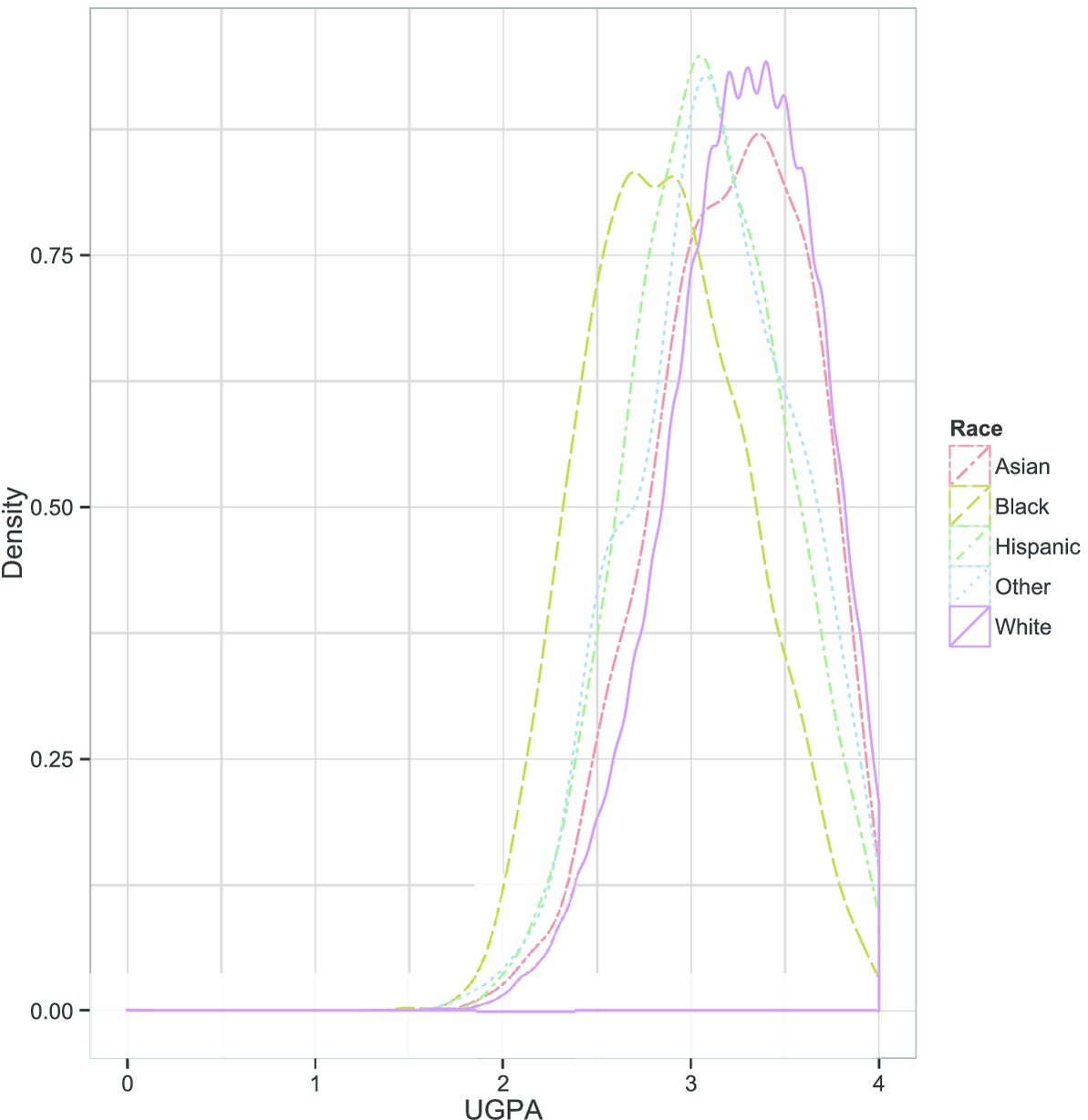}

\caption{UGPA distributions by racial group.
Although there is a substantial discrepancy between the white and black
student distributions, the gap is less pronounced for the other
minorities. Asian and white students seem especially aligned, while
Hispanic and Other students are aligned.}\label{ugparace}
\end{figure*}

We limited our analysis to white and black students due to
complications in characterizing AA policies toward Hispanic, Asian, and
``Other'' students. Specifically, as can be seen in Figures~\ref
{lsatrace}--\ref{ugparace}, Asian and ``Other'' students look very
similar to white students in terms of their LSAT scores and UGPAs, so
it is questionable whether they receive admissions preferences through
affirmative action. Also, although Hispanics generally are thought to
be treated similarly to black students in admissions preferences, we
found the trends in this dataset to be much less clear. For example, as
shown in Table~\ref{racecoeff}, the regression coefficients for the
probability of enrolling in each tier for Hispanic students do not
follow a specific trend. These factors make constructing SES-equivalent
categories for these racial groups dubious. Thus, for this analysis, we
focused our attention on examining the impact of the racial vs. SES AA
systems on black and white students.

%

%

\subsection{Calculating the SES Score}\label{appscore}

The SES score was computed as the first principal component of the SES
factors. The resulting score was the following:
\begin{eqnarray*}
\mathit{SES\_Score} &=& 0.442 \mathit{occ_{mom}} + 0.458 \mathit{occ_{dad}}
\\
&&{} + 0.485 \mathit{ed_{mom}}
+ 0.492 \mathit{ed_{dad}}
\\
&&{} + 0.342 \mathit{fam\_inc},
\end{eqnarray*}
where $\mathit{occ}$ is the parent's occupation category, $\mathit{ed}$ is the parent's
educational attainment, and $\mathit{fam\_inc}$ is the response to the general
SES question. All of the SES factors were on a scale from 1 to 5, with
higher numbers corresponding to higher SES, and were standardized
before calculating the principal component.

For 14,291 students, one or more components of their SES scores was
missing. Because this number of students is sufficiently large that
simply removing the students from the data would compromise the sample
size of minority students, we used the following method to impute the
missing SES data. For instances where $\mathit{occ}$ for a parent was missing
but $\mathit{ed}$ was available, we imputed the $\mathit{occ}$ of the parent as the $\mathit{ed}$
of the parent, and vice versa when $\mathit{ed}$ was missing for a parent.
Similarly, although a response of ``homemaker'' is technically not
missing data, it does not have a clear SES ranking relative to other
occupations; we replaced the $\mathit{occ}$ for homemakers\vadjust{\goodbreak} with the value of
their $\mathit{ed}$ to better capture their earning potentials. If $\mathit{occ}$ and
$\mathit{ed}$ were both missing for a parent, we imputed them with the $\mathit{occ}$ and
$\mathit{ed}$ of the other parent under the assumption that people tend to marry
within the same~SES.

If the information for both parents was missing for a student, we
assigned to the student the SES score corresponding to her $\mathit{fam\_inc}$
percentile rank. For example, if her $\mathit{fam\_inc}$ were 5, and if 80\% of
students had a $\mathit{fam\_inc}$ less than 5, we would impute her ``parental
score'' (the part of the score excluding $\mathit{fam\_inc}$) as the 80th
percentile among all parental scores. Thus, the student's relative
score would be similar to what it would be if the ranking system were
exclusively based on $\mathit{fam\_inc}$. Analogously, in cases where $\mathit{fam\_inc}$
was missing for a student, we calculated the percentile of the
student's parental score and imputed the student's $\mathit{fam\_inc}$ as the
$\mathit{fam\_inc}$ corresponding to that percentile.\vadjust{\goodbreak} We removed students with
no SES information available.

%
\begin{table*}
\tabcolsep=0pt
\caption{Regression coefficients from logistic regression for all races}\label{racecoeff}
\begin{tabular*}{\tablewidth}{@{\extracolsep{\fill}}@{}ld{3.2}d{2.2}d{2.2}d{2.2}d{2.2}d{2.2}@{}}
\hline
\textbf{Tier} & \multicolumn{1}{c}{\textbf{Intercept}} & \multicolumn{1}{c}{\textbf{LSAT}} & \multicolumn{1}{c}{\textbf{Asian}}
& \multicolumn{1}{c}{\textbf{Black}} & \multicolumn{1}{c}{\textbf{Hispanic}} & \multicolumn{1}{c@{}}{\textbf{Other}}\\
\hline
1 &-12.37 & 0.25 & 0.84 & 1.84 & 0.36 & 0.59 \\
2 & -5.64 & 0.11 & 0.89 & 0.92 & -0.28 & 0.63 \\
3 & -2.78 & 0.05 & -0.32 & 0.73 & 1.24 & -0.39 \\
4 & 3.80 &-0.12 & -0.50 &-1.31 & -1.48 & -0.11 \\
5 & 5.24 &-0.21 & -1.44 &-2.25 & 0.84 & -1.16 \\
\hline
\end{tabular*}
\tabnotetext[]{ta3}{\textit{Note}:
The results in this table were derived from logistic regressions with
an indicator for being in a given tier as the outcome variable and race
and LSAT as predictors. They show that effects of affirmative action in
admissions are more apparent when comparing Black and White students
than other minorities. What was estimated was the enrollment
probability, not the enrollment probability conditional on not having
been enrolled in a higher tier. Consequently, racial groups benefitting
from affirmative action should have positive race coefficients for
upper tiers and negative race coefficients for lower tiers. The results
for Hispanic students are thus not very interpretable. Moreover,
although the coefficients for Asian and Other follow the expected
trends, their elevated race coefficients for higher tiers are
surprising given that Asian and Other students generally do not benefit
from affirmative action. These trends might be the product of Asian and
Other students coming from better undergraduate institutions or having
better extracurricular records.}\vspace*{15pt}
\end{table*}

%
\begin{figure*}[b]

\includegraphics{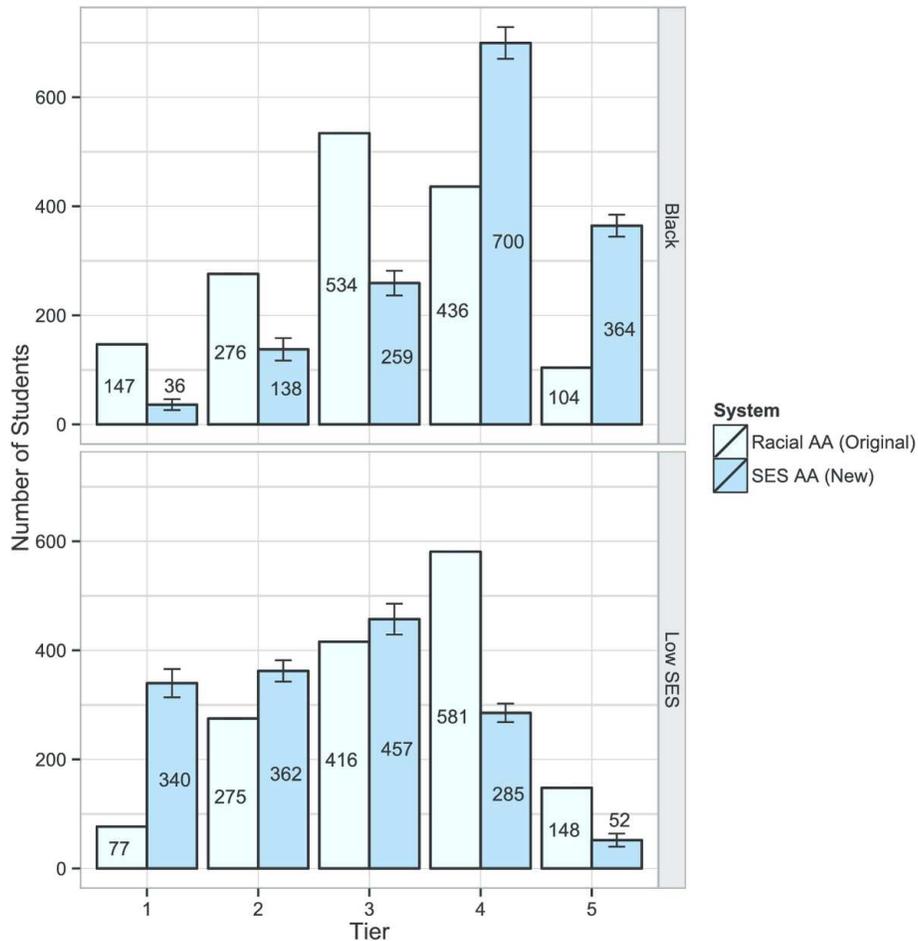}

\caption{Unconstrained model, impact of SES AA on demographic
composition across tiers.
In the unconstrained model, students are enrolled into each tier
(starting with Tier~1 and going down to Tier~5) until each tier is
filled, but without constraints on the number of low~SES students in
each tier. This figure shows that if SES students are given the same AA
boost that black students received without constraints on~the numbers
of low SES students enrolled into higher tiers, they experience
disproportionate increases in their enrollment in higher tiers.}\label{figdemcompcon}
\end{figure*}

%
\begin{figure*}\vspace*{-3pt}
\includegraphics{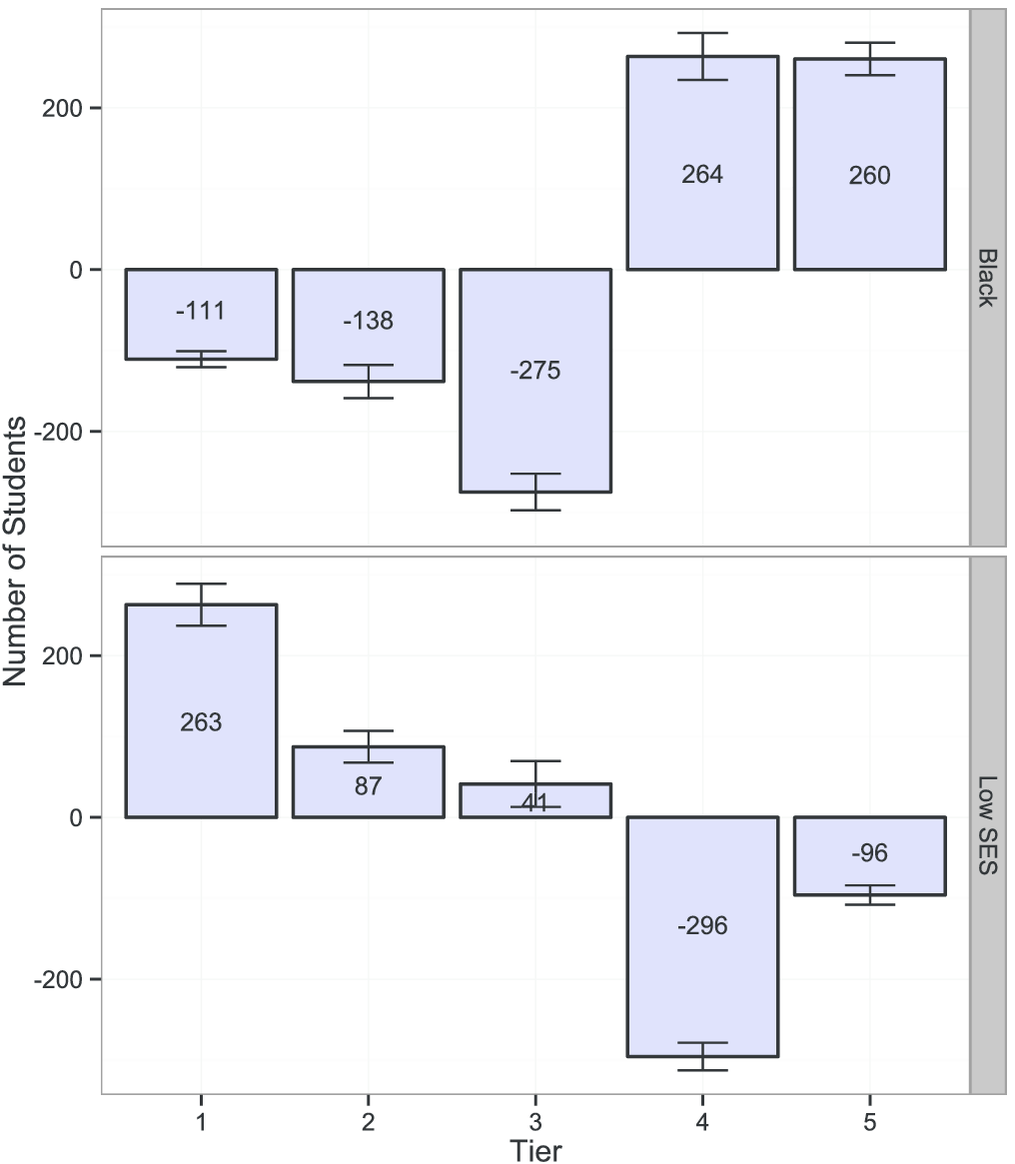}
\vspace*{-5pt}
\caption{Unconstrained model, impact of SES AA on numbers of
AA-targeted students across tiers.
This figure illustrates that the increase in low SES students in Tiers
1--3 is far greater than the decrease in black students in these tiers,
suggesting that if low SES students simply received the same boost as
black students without constraints, their numbers in higher tiers would
increase disproportionately. This is a result of the larger achievement
gap between black and white students than between low SES and high
SES students.}\label{figdemcompdiffcon}\vspace*{-8pt}
\end{figure*}

\subsection{Undergraduate GPA (UGPA)}\label{appugpa}

Although the data included the students' UGPAs, it did not include any
information about the undergraduate institutions the students attended,
thus rendering UGPA less interpretable. In carrying out our analysis,
we included UGPA as a predictor for enrollment probabilities and
academic outcomes, but the coefficients for UGPA should be interpreted
carefully given this ambiguity. In general, LSAT is the more reliable
metric for student academic ability given that it is standardized for
all students. Thus, throughout the paper, we often examine changes in
LSAT distribution in order to gauge changes in relative academic
ability across law school tiers and demographic groups.\vadjust{\goodbreak}

\subsection{Diversity Quotas}\label{appquota}
The rationale for modeling the admissions process with diversity quotas
is that admissions committees are presumably less concerned with
maintaining the size of the boost they give to AA-targeted students and
more concerned with the outcomes---the numbers of AA-targeted students
who enroll. Without this restriction, low SES students under SES AA
would functionally receive the same increases in enrollment probability
that black students received under the race-based AA system. This is
problematic because low SES students in general have higher LSAT scores
than black students (Figure~\ref{lsatrace}), so simply using the black
student enrollment functions for low SES students would result in
excessive numbers of low SES students entering top tiers, as shown in
Figures~\ref{figdemcompcon}--\ref{figdemcompdiffcon}. For example, the
unconstrained model predicts the number of low SES students in Tier~1
increases from 77 to~340.\vadjust{\goodbreak}

Although disproportionate increases in low SES student enrollment are
conceivable, it is questionable whether admissions committees would be
willing to allocate many more slots to low SES students under SES AA.
Given that admissions committees exercise affirmative action with the
goal of achieving a diverse incoming class, and given that a smaller
boost would suffice to yield a socioeconomically diverse student body,
it is more likely that admissions committees would offer low SES
students a less substantial boost than the one currently offered to
black students. Using a quota model thus reflects this mitigation of
the size of the admissions boost.

\subsection{No-Quota Model}\vspace*{-10pt}\label{appnoquota}

\newpage

\section{Algorithms}\label{appalg}

\subsection{Estimating Tier~Enrollment Probability Functions}\label{appenroll}
\begin{longlist}[7.]
\item[1.] Create separate lists for black and white students.
\item[2.] Run the logistic regression for the black and white students
separately, with an indicator random variable of being in Tier~1 (vs.
Tiers 2--5) as the outcome variable.
\item[3.] Remove the Tier~1 students from the black and white student lists.
\item[4.] Run the logistic regression for the remaining black and white
students, with an indicator random variable of being in Tier~2 (vs.
Tiers 3--5) as the outcome variable.
\item[5.] Remove the Tier~2 students from both lists.
\item[6.] Repeat for Tiers 3 and 4.
\item[7.] The conditional enrollment probability for Tier~5 is 1 for both
black and white students.\vadjust{\goodbreak}
\end{longlist}

\subsection{Reassigning Tiers for SES AA System}\label{appreassign}
\begin{enumerate}
\item[1.] Starting with Tier~1, draw the parameters for the probability
enrollment functions ($p_{i,1}^b$ and $p_{i,1}^w$) from the estimated
posterior distribution.
\item[2.] Separate the list of students into low and high SES.
\item[3.] Take a weighted sample of the low SES students drawing $N_{t}^b$
of them with probability weights $p_{i,1}^b$, where $N_{b, t}$ is the
number of black students in Tier~$t$ under racial AA and $p_{i,1}^b$
$p_{i,1}^b$ is the\vspace*{1pt} probability for student $i$ enrolling in Tier~1 as
a black student.\looseness=1
\item[4.] Perform the same procedure for the high SES students, drawing
$N_{t}^w$ of them with probability weights~$p_{i,1}^w$.
\item[5.] Take out the low and high SES students assigned to Tier~1 from
their respective lists, so they will not be eligible for reassignment
to lower tiers.
\item[6.] Repeat for Tiers 2--4, going from the most to least selective.
\item[7.] Assign remaining students to Tier~5.
\end{enumerate}

\subsection{Imputing Academic Outcomes}\label{appacout}
\begin{enumerate}
\item[1.]{\emph{Dropout}}: For each tier, use logistic regression
to find a function for student dropout probability based on the
original data.
\item[2.] Recalculate LSAT percentiles for each student based on their SES
AA tier assignment.
\item[3.] Draw parameters from the posterior distribution of the fit. Use
these parameters to impute the new dropout probabilities $d_{i,t}$ for
students by applying the function corresponding to their assigned law
school tier under SES AA, where $d_{i,t}$ is the probability that
student $i$ would dropout after attending Tier~$t$.
\item[4.] For each tier, go through the list of students once and draw from
a Bernoulli random variable with probability $d_{i,t}$ for each student
to impute whether they did or did not drop out under the SES AA system.
\item[5.]{\emph{Take Bar}}: Considering the list of students who
graduated under the original racial system, use logistic regression to
find functions, for each tier, of the probability of a student deciding
not to take the bar exam.
\item[6.] Now consider the set of students who were imputed to have
graduated from law school under the SES AA system. Use the function
corresponding to their newly assigned law school tier to impute their
probabilities of taking the bar.
\item[7.] Impute the outcome of taking the bar or not by going through the
list of students and drawing from a Bernoulli random variable with the
estimated probabilities of taking the bar exam.
\item[8.]{\emph{Bar Passage}}: For the students whose outcomes are
that they would take the bar, impute whether they pass the first time
using the same logistic regression method.
\item[9.] For the remaining students who did not pass the bar the first
time, impute whether they eventually pass the bar or fail using the
same basic method.
\end{enumerate}

\newpage

\section{Results}\label{appresults}

\setcounter{table}{3}
%
\begin{table}[h]
\tabcolsep=0pt
\caption{Overall changes in academic outcomes}\label{overallchange}
\begin{tabular*}{\tablewidth}{@{\extracolsep{\fill}}@{}lcc@{}}
\hline
\textbf{Outcome} & \textbf{Original} & \textbf{New (SE)} \\
\hline
Dropout rate & 0.0926 & 0.0937 (0.0036) \\
Rate of not taking bar & 0.0633 & 0.0624 (0.0030) \\
First-try bar passage rate & 0.752\phantom{0} & 0.753 (0.0053)\phantom{0} \\
Later-try bar passage rate & 0.0507 & 0.0508 (0.0025) \\
Bar failure rate & 0.0393 & 0.0399 (0.0024) \\
\hline
\end{tabular*}
\tabnotetext[]{ta4}{\textit{Note}:
None of the academic outcomes change substantially in aggregate between
the two systems.}\vspace*{15pt}
%
%
\tabcolsep=0pt
\caption{Percentage changes in outcomes by race}\label{tabchangerace}
\begin{tabular*}{\tablewidth}{@{\extracolsep{\fill}}@{}lcd{2.4}@{}}
\hline
\textbf{Outcome variable} & \textbf{Race} & \multicolumn{1}{c@{}}{\textbf{Percentage change}}\\
\hline
Dropout rate & Black & 0.1196 \\
& White & -0.0023 \\[2pt]
Did not attempt bar & Black & 0.0755 \\
& White & 0.0081 \\[2pt]
Passed bar first try & Black & -0.0719 \\
& White & -0.0009 \\[2pt]
Passed bar later try & Black & -0.0622 \\
& White & 0.0108 \\[2pt]
Failed to pass bar & Black & 0.0812 \\
& White & -0.0031 \\
\hline
\end{tabular*}
\tabnotetext[]{ta5}{\textit{Note}:
The quantities are expressed as percentage change in proportions in
going from the racial to SES AA system.}\vspace*{15pt}
%
%
\tabcolsep=0pt
\caption{Percentage changes in outcomes by SES}\label{tabchangeses}
\begin{tabular*}{\tablewidth}{@{\extracolsep{\fill}}@{}lcd{2.4}@{}}
\hline
\textbf{Outcome variable} & \textbf{SES} & \multicolumn{1}{c@{}}{\textbf{Percentage change}}\\
\hline
Dropout rate & Low SES & -0.0051 \\
& High SES & 0.0137 \\[2pt]
Did not attempt bar & Low SES & 0.0116 \\
& High SES & 0.0129 \\[2pt]
Passed bar first try & Low SES & -0.0172 \\
& High SES & -0.0027 \\[2pt]
Passed bar later try & Low SES & 0.0795 \\
& High SES & -0.0041 \\[2pt]
Failed to pass bar & Low SES & 0.0540 \\
& High SES & 0.0109 \\
\hline
\end{tabular*}
\tabnotetext[]{ta6}{\textit{Note}:
The quantities are expressed as percentage change in proportions in
going from the racial to SES AA system.}
\end{table}

\setcounter{figure}{13}
\begin{figure*}[t!]
\vspace*{100pt}
\includegraphics{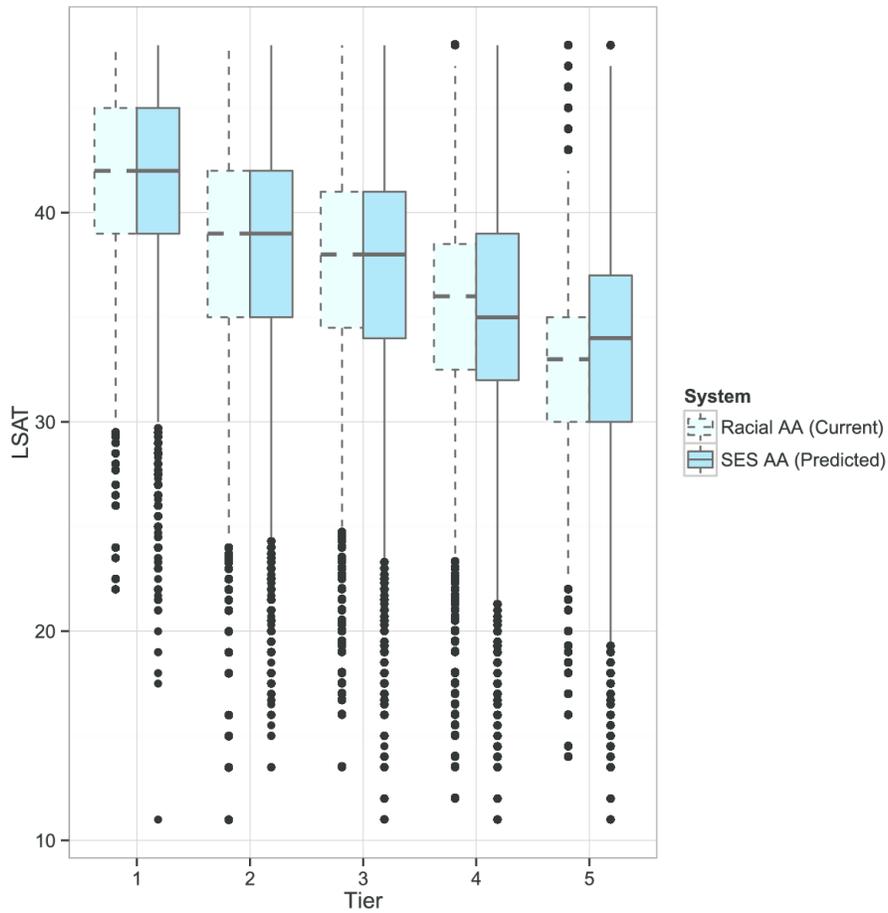}

\caption{Impact of SES AA on LSAT distribution by tier.
LSAT distributions remain roughly the same between racial and SES AA
for Tiers 1--3. The distributions for Tiers 4 and 5 widen, with Tier~4
having a lower mean and Tier~5 a higher mean. Thus, the academic
qualifications in each tier are not changing substantially between the
two systems even as their demographic compositions change.}\label{lsatboxtier}
\end{figure*}

\newpage
\mbox{}

\newpage
%
\setcounter{figure}{14}
\begin{figure*}[t!]
\vspace*{100pt}
\includegraphics{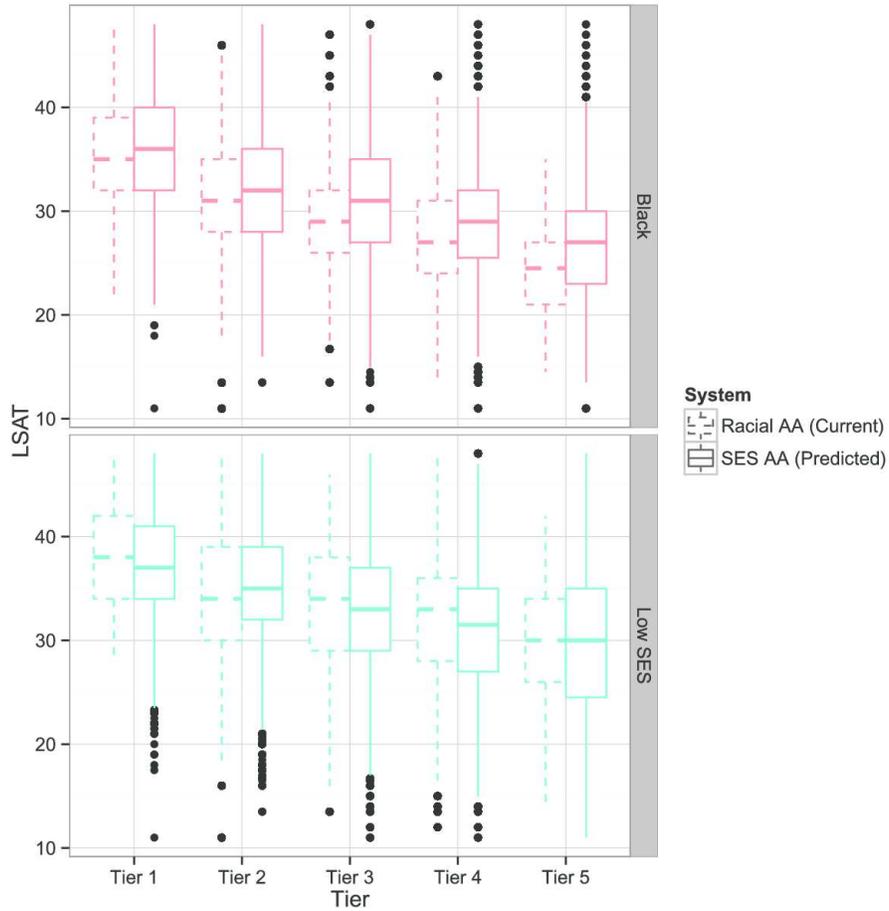}

\caption{Impact of SES AA on LSAT distribution by tier for black and
low SES students.
In going from racial AA to SES AA, LSAT distributions are shifted
toward higher LSAT scores for black students in all tiers and toward
lower LSAT scores for low SES students for Tiers 1, 3, and 4. Given
that the overall LSAT distributions have not changed substantially, as
shown in Figure~\protect\ref{lsatboxtier}, this suggests that black students
are better academically matched to their tiers under SES AA than under
racial AA, whereas low SES students are better academically matched to
their tiers under racial AA than under SES AA. Thus, by analyzing the
simulation results for SES AA, we can gauge whether there is evidence
for a mismatch effect.}\label{lsatboxtiergroup}
\end{figure*}
%

\newpage

\subsection{Changes in LSAT Distribution}\label{applsat}
\vspace{-20pt}

\setcounter{figure}{15}
%
\begin{figure*}[t!]
\vspace*{100pt}
\includegraphics{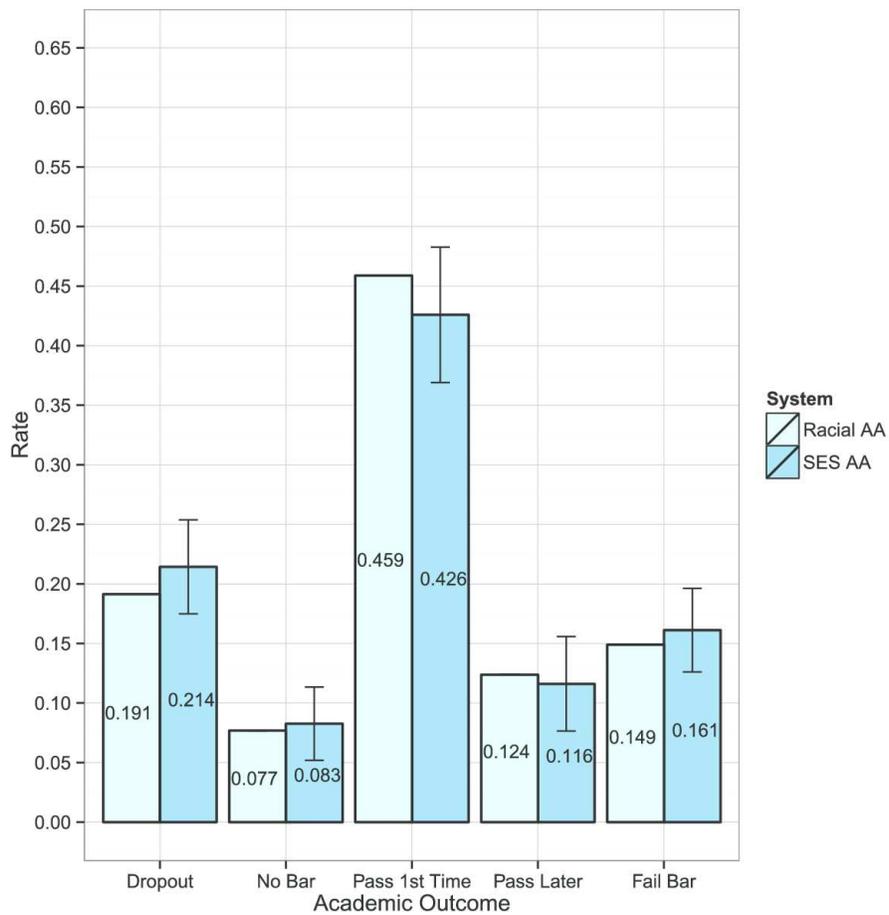}

\caption{Impact of SES AA on graduation and bar passage outcomes for
black students.
Academic outcomes seem to worsen or stay the same overall for black
students under the SES AA system, a result that directly contradicts
the predictions of the mismatch hypothesis.}\label{blackacfig}
\end{figure*}

\begin{figure*}[t!]
\vspace*{100pt}
\includegraphics{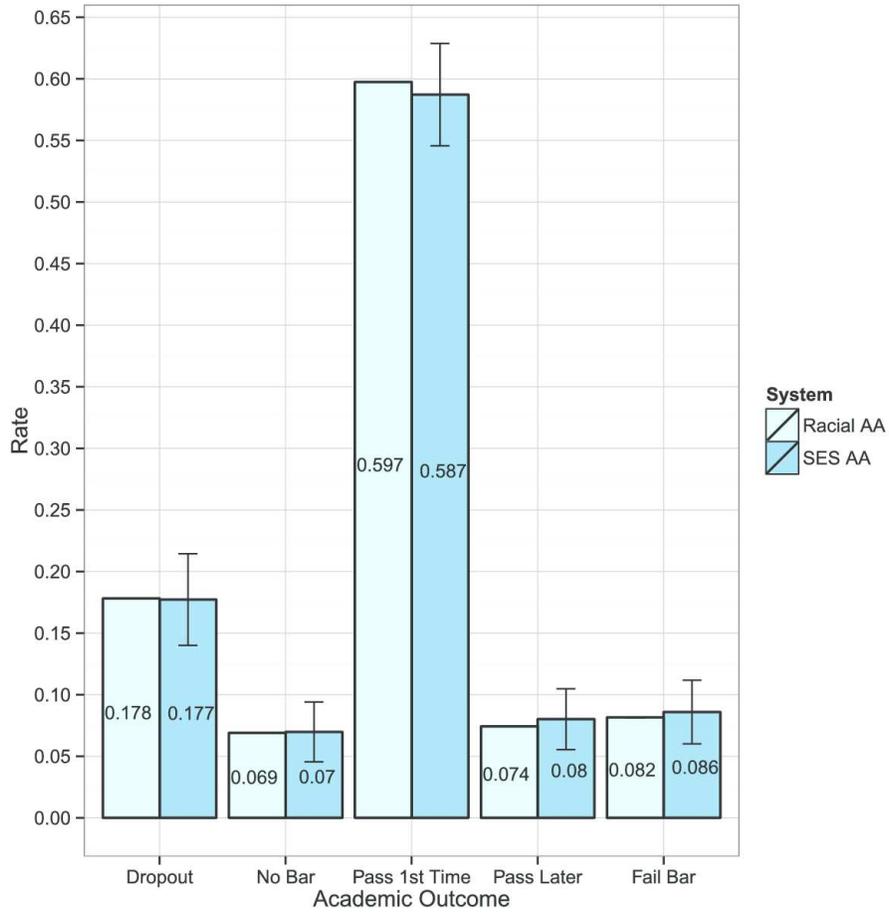}

\caption{Impact of SES AA on graduation and bar passage outcomes for
low SES students.
Academic outcomes seem to improve or stay the same overall for low SES
students under the SES AA system, a result that directly contradicts
the predictions of the mismatch hypothesis.}\label{lowsesacfig}
\end{figure*}

\newpage

\mbox{}

\newpage

\newpage

\mbox{}

\newpage

\subsection{Changes by Demographic Group}\label{appdem}
\vspace{100pt}

\newpage

\mbox{}

\newpage

\newpage

\mbox{}

\newpage

\setcounter{figure}{17}
%
\begin{figure*}[t!]
\vspace*{100pt}
\includegraphics{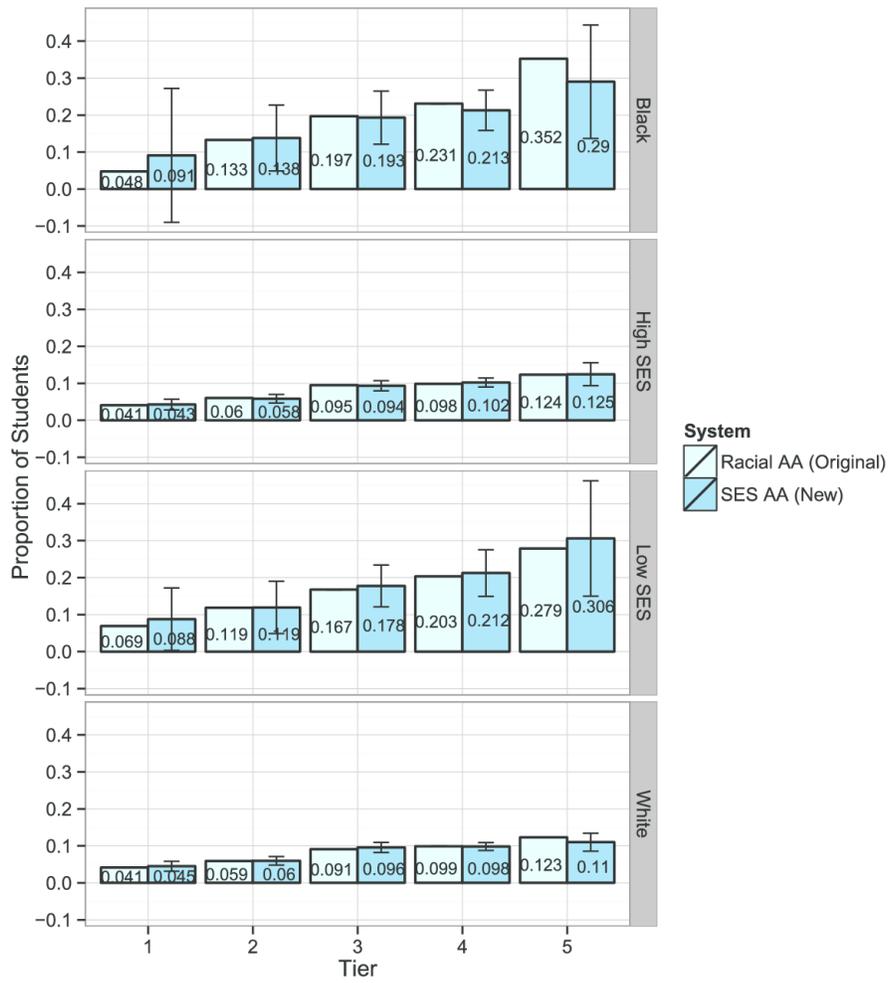}

\caption{Impact of SES AA on dropout rates.
There are no statistically significant changes in dropout rates in
going from racial AA to SES AA. Note that there were only seven black
students who dropped out in the actual data, contributing to the large
standard errors for that prediction.}\label{dropouttierfig}
\end{figure*}

\newpage

\mbox{}

\newpage

\subsection{Changes by Tier}\label{apptier}
\vspace*{-20pt}

%
%

%
\begin{figure*}[t!]
\vspace*{100pt}
\includegraphics{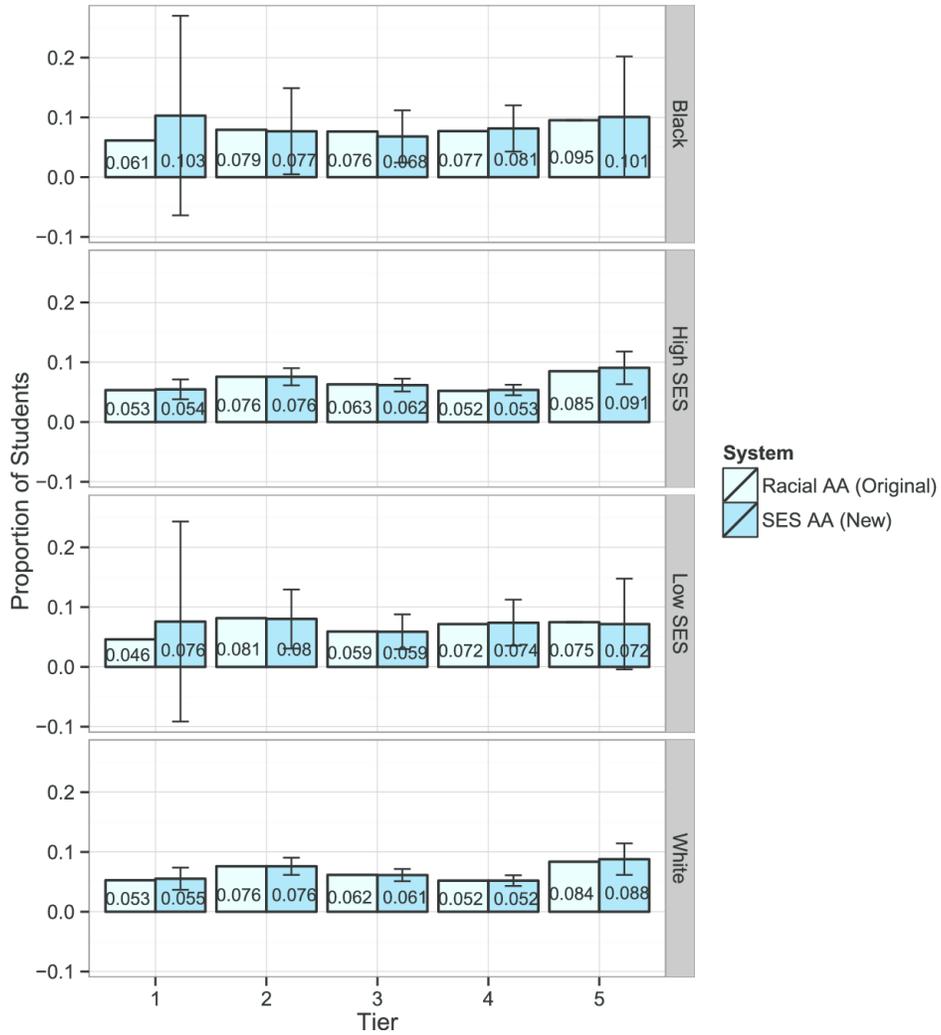}

\caption{Impact of SES AA on rates of not taking the bar.
Rates of not taking the bar increase for black students in Tier~1 and
Tiers 4--5 but decrease for Tiers 2--3. The rates generally decrease for
low SES students.}\label{nobartierfig}
\end{figure*}

\newpage

\mbox{}

\newpage

%
\begin{figure*}[t!]
\vspace*{100pt}
\includegraphics{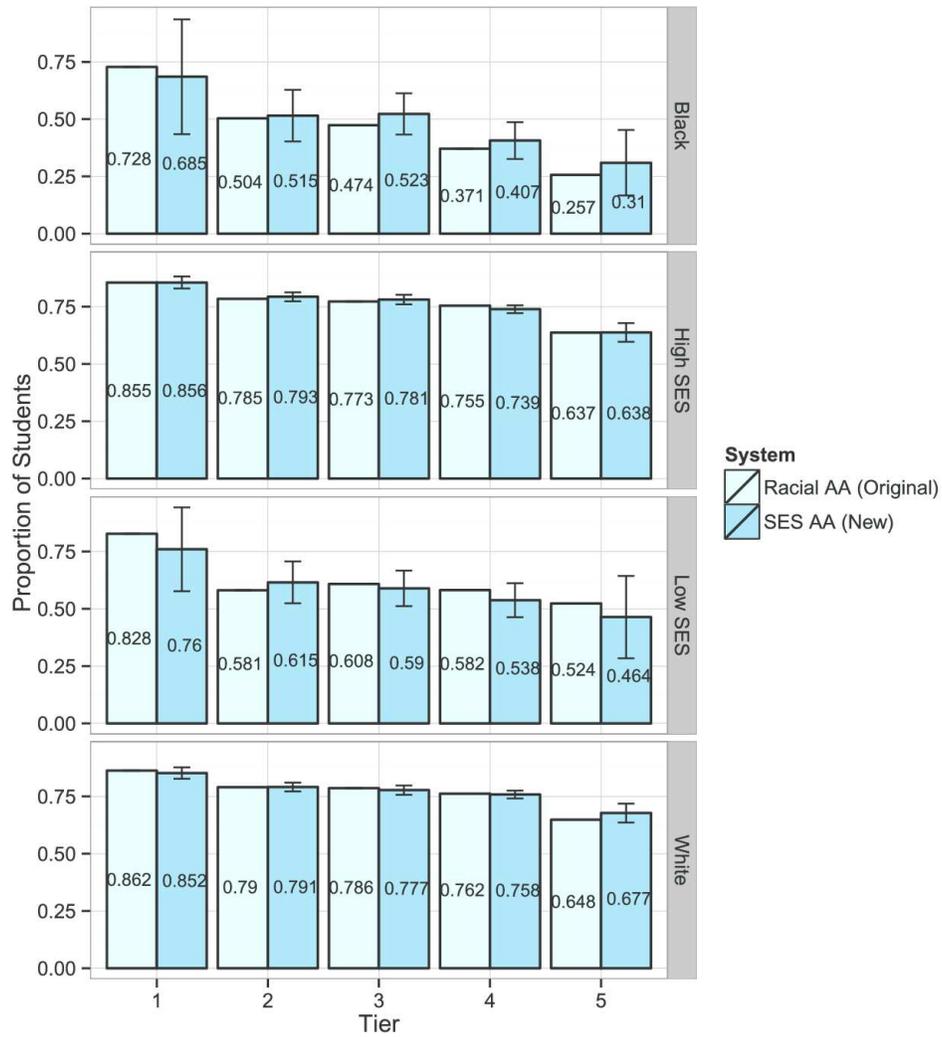}

\caption{Impact of SES AA on rates of passing bar on first try.
First-try bar passage rates decrease for black students in Tier~1 but
increase for lower tiers. For low SES students, they increase for Tiers
1--3 but decrease for Tiers 4--5.}\label{passfirsttierfig}
\end{figure*}

\newpage

\mbox{}

\newpage

\newpage

\mbox{}

\newpage

%
\begin{figure*}[t!]
\vspace*{100pt}
\includegraphics{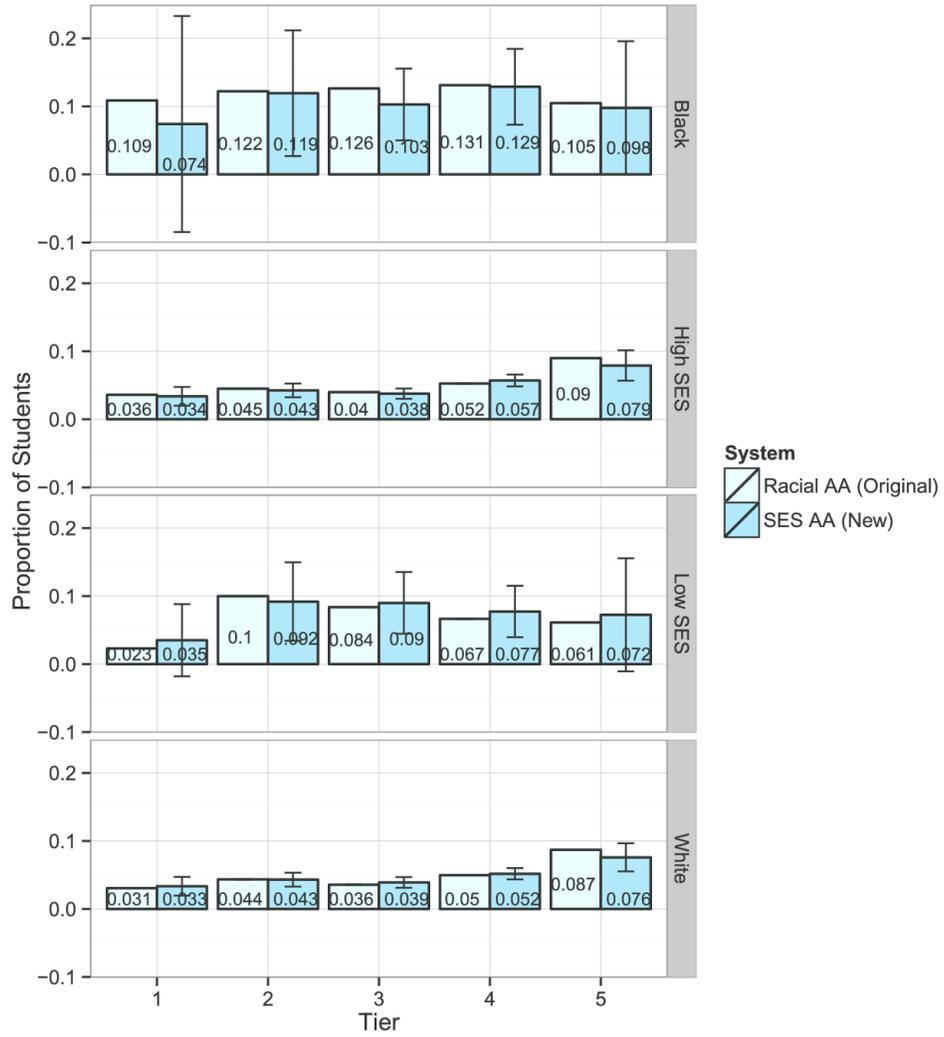}

\caption{Impact of SES AA on rates of passing bar on later try.
Later-try bar passage rates generally decrease for black students. The
rates remain fairly constant for the other demographic groups.}\label{passlatertierfig}
\end{figure*}

\newpage

\mbox{}

\newpage

\newpage

\mbox{}

\newpage

%
\begin{figure*}[t!]
\vspace*{100pt}
\includegraphics{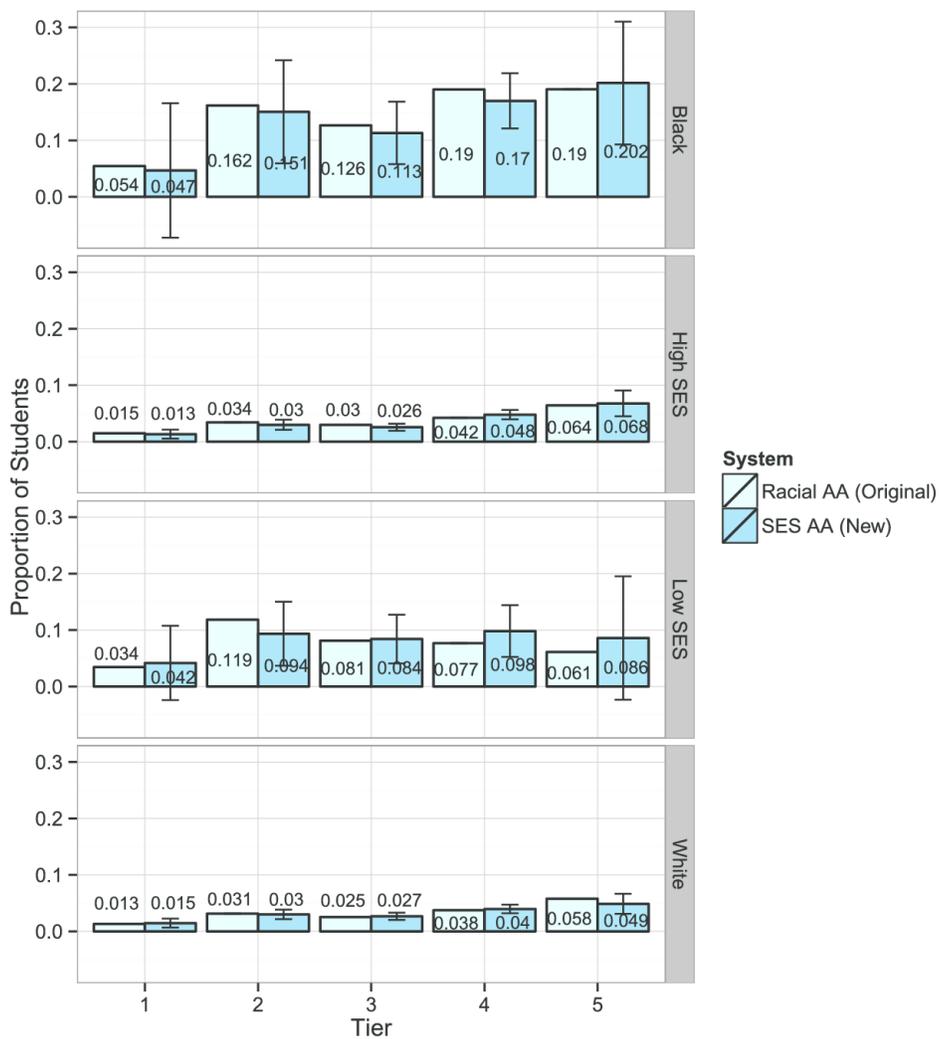}

\caption{Impact of SES AA on bar failure rates.
Bar failure rates increase for black students in Tier~1 and decrease
across the remainder of the tiers, though with large error bars. The
rates increase slightly for low SES students in Tier~1 and Tiers 4--5
but decrease for Tiers 2--3.}\label{failtierfig}
\end{figure*}

\newpage

\mbox{}

\newpage

\newpage

\mbox{}

\newpage

\newpage

\mbox{}

\newpage

\begin{table*}[t!]
\vspace*{100pt}
\hspace*{-2pt}\centerline{
\begin{minipage}{\columnwidth}
\setcounter{table}{6}
%
\begin{table}[H]
\tabcolsep=0pt
\caption{Fitted values: Academic outcomes by race}\label{fitrace}
\begin{tabular*}{\tablewidth}{@{\extracolsep{\fill}}@{}lccc@{}}
\hline
\textbf{Outcome variable} & & \textbf{Original} & \textbf{Fitted (SE)} \\
\hline
Dropout rate & Black & 0.1914 & 0.1922 (0.0177) \\
& White & 0.0866 & 0.0868 (0.0029) \\[2pt]
Did not attempt bar & Black & 0.0768 & 0.0837 (0.0138) \\
& White & 0.0617 & 0.0625 (0.0024) \\[2pt]
Passed bar first try & Black & 0.4589 & 0.4490 (0.0233) \\
& White & 0.7731 & 0.7722 (0.0046) \\[2pt]
Passed bar later try & Black & 0.1238 & 0.1219 (0.0123) \\
& White & 0.0461 & 0.0463 (0.0024) \\[2pt]
Failed to pass bar & Black & 0.1490 & 0.1532 (0.0161) \\
& White & 0.0326 & 0.0322 (0.0019) \\
\hline
\end{tabular*}
\end{table}
\end{minipage}\hspace*{21pt}
\begin{minipage}{\columnwidth}
\setcounter{table}{7}
%
%
\begin{table}[H]
\caption{Fitted values: Academic outcomes by SES}\label{fitses}
\begin{tabular*}{\tablewidth}{@{\extracolsep{\fill}}@{}lccc@{}}
\hline
\textbf{Outcome variable} & & \textbf{Original} & \textbf{Fitted (SE)} \\
\hline
Dropout rate & Low SES & 0.1781 & 0.1787 (0.0162) \\
& High SES & 0.0874 & 0.0877 (0.0030) \\[2pt]
Did not attempt bar & Low SES & 0.0689 & 0.0730 (0.0133) \\
& High SES & 0.0621 & 0.0631 (0.0024) \\[2pt]
Passed bar first try & Low SES & 0.5974 & 0.5975 (0.0224) \\
& High SES & 0.7646 & 0.7631 (0.0047) \\[2pt]
Passed bar later try & Low SES & 0.0742 & 0.0713 (0.0128) \\
& High SES & 0.0491 & 0.0494 (0.0022) \\[2pt]
Failed to pass bar & Low SES & 0.0815 & 0.0794 (0.0127) \\
& High SES & 0.0367 & 0.0367 (0.0018) \\
\hline
\end{tabular*}
\end{table}
\end{minipage}}
\end{table*}

\newpage

\mbox{}

\newpage

\section{Model Fit for Academic Outcomes}\label{appfit}

We simulated the academic outcomes based on the actual tier assignments
and found that the model successfully predicted all of the quantities
of interest to within a 95\% confidence interval.\vspace*{20pt}

\setcounter{table}{8}
%
\begin{table*}[t!]
\tabcolsep=0pt
\caption{Regression coefficients for dropout rates}\label{regdropout}
\begin{tabular*}{\tablewidth}{@{\extracolsep{\fill}}@{}ld{2.8}d{2.8}d{2.8}d{2.8}d{2.8}@{\hspace*{-2pt}}}
\hline
\textbf{Coefficient} & \multicolumn{1}{c}{\textbf{Tier~1 (SE)}} & \multicolumn{1}{c}{\textbf{Tier~2 (SE)}}
& \multicolumn{1}{c}{\textbf{Tier~3 (SE)}} & \multicolumn{1}{c}{\textbf{Tier~4 (SE)}} &
\multicolumn{1}{c@{}}{\textbf{Tier~5 (SE)}} \\
\hline
Intercept & 5.06~(2.37) & 0.77~(1.22) & 0.91~(0.77) & -1.09~(0.73) & 1.44~(1.34) \\
Female & -0.62~(0.22) & -0.14~(0.12) & -0.06~(0.08) & -0.07~(0.07) &-0.25~(0.14) \\
LSAT & -0.19~(0.07) & -0.08~(0.04) & -0.09~(0.02) & -0.03~(0.02) &-0.09~(0.05) \\
LSAT percentile & 1.74~(1.11) & -0.08~(0.71) & 0.37~(0.44) & -0.43~(0.40) & 0.18~(0.72) \\
UGPA & -0.22~(0.30) & -0.16~(0.16) & -0.01~(0.11) & 0.08~(0.09) & -0.15~(0.16) \\
Black & -1.56~(0.59) & 0.21~(0.28) & 0.31~(0.19) & 0.71~(0.18) & 0.71~(0.34) \\
Low SES & 0.64~(0.72) & 0.34~(0.30) & 0.41~(0.21) & 0.15~(0.18) & 0.64~(0.33) \\
LSAT percentile: Black & 4.79~(1.32) & 0.32~(1.13) & -0.96~(1.00) &-0.79~(0.89) & -1.15~(1.79) \\
LSAT percentile: Low SES & -4.91~(4.32) & 0.16~(0.80) & 0.26~(0.52) &1.09~(0.39) & 0.49~(0.68) \\
\hline
\end{tabular*}\vspace*{10pt}
%
%
\tabcolsep=0pt
\caption{Regression coefficients for rates of not taking the bar}\label{regnobar}
\begin{tabular*}{\tablewidth}{@{\extracolsep{\fill}}@{}ld{3.9}d{2.8}d{2.8}d{2.8}d{2.8}@{\hspace*{-2pt}}}
\hline
\textbf{Coefficient} & \multicolumn{1}{c}{\textbf{Tier~1 (SE)}} & \multicolumn{1}{c}{\textbf{Tier~2 (SE)}}
& \multicolumn{1}{c}{\textbf{Tier~3 (SE)}} & \multicolumn{1}{c}{\textbf{Tier~4 (SE)}} &
\multicolumn{1}{c@{}}{\textbf{Tier~5 (SE)}} \\
\hline
Intercept & -2.98~( 3.00) & -0.56~(1.40) & 2.26~(1.05) & -3.09~(1.12) &-3.45~(1.74) \\
Female & -0.17~( 0.19) & -0.06~(0.11) & 0.08~(0.10) & 0.07~(0.09) &0.04~(0.16) \\
LSAT & -0.12~( 0.08) & -0.04~(0.04) & -0.16~(0.03) & 0.01~(0.04) & 0.01~(0.06) \\
LSAT percentile & 2.61~( 1.22) & 0.80~(0.79) & 2.41~(0.58) & -0.18~(0.60) & 0.23~(0.97) \\
UGPA & 1.13~( 0.33) & -0.18~(0.15) & -0.02~(0.13) & -0.01~(0.12) & 0.25~(0.19) \\
Black & 0.54~( 0.57) & 0.44~(0.35) & -0.22~(0.26) & 0.50~(0.27) & 1.25~(0.51) \\
Low SES & 1.91~( 0.78) & 0.27~(0.34) & -0.42~(0.33) & 0.40~(0.27) &-1.11~(0.70) \\
LSAT percentile: Black & -0.05~( 1.85) & -3.36~(1.92) & 0.78~(0.91) &0.16~(0.98) & -4.34~(3.59) \\
LSAT percentile: Low SES & -19.35~(10.27) & -0.67~(0.89) & 0.79~(0.74) & 0.05~(0.59) & 2.36~(1.10) \\
\hline
\end{tabular*}
\end{table*}
%
%
\begin{table*}[t!]
\tabcolsep=0pt
\caption{Regression coefficients for rates of passing bar on first try}\label{regfirst}
\begin{tabular*}{\tablewidth}{@{\extracolsep{\fill}}@{}ld{2.8}d{2.8}d{2.8}d{2.8}d{2.8}@{\hspace*{-2pt}}}
\hline
\textbf{Coefficient} & \multicolumn{1}{c}{\textbf{Tier~1 (SE)}} & \multicolumn{1}{c}{\textbf{Tier~2 (SE)}}
& \multicolumn{1}{c}{\textbf{Tier~3 (SE)}} & \multicolumn{1}{c}{\textbf{Tier~4 (SE)}} &
\multicolumn{1}{c@{}}{\textbf{Tier~5 (SE)}} \\
\hline
Intercept & -7.79~(2.12) & -7.16~(1.25) & -7.61~(0.94) & -6.51~(0.79) & -7.71~(1.72) \\
Female & 0.07~(0.20) & -0.22~(0.12) & -0.08~(0.10) & -0.12~(0.08) & -0.48~(0.13) \\
LSAT & 0.19~(0.06) & 0.18~(0.04) & 0.19~(0.03) & 0.17~(0.03) & 0.25~(0.06) \\
LSAT percentile & -0.43~(1.05) & -0.59~(0.76) & -0.57~(0.55) & -0.47~(0.44) & -2.09~(0.90) \\
UGPA & 0.94~(0.27) & 0.99~(0.14) & 1.04~(0.12) & 0.96~(0.10) & 0.76~(0.17) \\
Black & 0.06~(0.37) & -0.36~(0.24) & -0.12~(0.19) & -0.69~(0.19) & -0.33~(0.41) \\
Low SES & 0.84~(0.65) & -0.76~(0.27) & -0.44~(0.23) & -0.09~(0.21) & 0.59~(0.46) \\
LSAT percentile: Black & -1.72~(1.51) & -0.83~(0.91) & -0.68~(0.83) & -0.13~(0.81) & -0.94~(1.51) \\
LSAT percentile: Low SES & -2.34~(1.57) & 0.29~(0.87) & 0.34~(0.79) & -0.01~(0.58) & -0.74~(1.03) \\
\hline
\end{tabular*}
\end{table*}

%
%
\begin{table*}
\tabcolsep=0pt
\caption{Regression coefficients for rates of passing bar on later try}\label{reglater}
\begin{tabular*}{\tablewidth}{@{\extracolsep{\fill}}@{}ld{2.8}d{2.8}d{2.8}d{2.8}d{2.8}@{\hspace*{-2pt}}}
\hline
\textbf{Coefficient} & \multicolumn{1}{c}{\textbf{Tier~1 (SE)}} & \multicolumn{1}{c}{\textbf{Tier~2 (SE)}}
& \multicolumn{1}{c}{\textbf{Tier~3 (SE)}} & \multicolumn{1}{c}{\textbf{Tier~4 (SE)}} &
\multicolumn{1}{c@{}}{\textbf{Tier~5 (SE)}} \\
\hline
Intercept & -4.61~(3.38) & -4.92~(1.85) & -2.52~(1.38) & -3.37~(1.18) & -4.41~(2.31) \\
Female & 0.27~(0.41) & 0.27~(0.21) & 0.10~(0.18) & -0.15~(0.14) & -0.11~(0.23) \\
LSAT & 0.12~(0.09) & 0.14~(0.06) & 0.04~(0.04) & 0.11~(0.04) & 0.15~(0.08) \\
LSAT percentile & -1.04~(1.88) & -1.94~(1.20) & -0.31~(0.90) & -1.78~(0.72) & -1.14~(1.26) \\
UGPA & 0.37~(0.56) & 0.30~(0.27) & 0.52~(0.22) & 0.27~(0.17) & 0.19~(0.30) \\
Black & 0.26~(0.66) & -0.21~(0.37) & 0.30~(0.31) & -0.47~(0.29) & 0.31~(0.60) \\
Low SES & -0.38~(1.15) & -0.64~(0.44) & -0.25~(0.36) & -0.31~(0.32) & 0.21~(0.78) \\
LSAT percentile: Black & -0.17~(3.20) & 0.30~(1.57) & -3.49~(1.87) & 1.32~(1.77) & -6.78~(4.99) \\
LSAT percentile: Low SES & -2.95~(3.45) & 2.56~(1.92) & 0.74~(1.70) & 0.83~(1.09) & -1.07~(1.99) \\
\hline
\end{tabular*}
\end{table*}

\newpage

\mbox{}

\newpage

\section{Regression Coefficients}\vspace*{5pt}\label{appreg}

\end{appendix}

\newpage

\mbox{}

\newpage

\section*{Acknowledgments}
We thank our peer reviewers for their thoughtful comments and
insightful suggestions. Also, a special thanks to Raj Chetty for
helpful suggestions to related literature and to the other members of
the Statistics and Economics Departments at Harvard who contributed
their advice and support to this research project.


%

%
\end{document}